\documentclass[aps,physrev,reprint,superscriptaddress,amsmath,amssymb,longbibliography,floatfix]{revtex4-2}
\usepackage{mathtools,booktabs,graphicx}

\usepackage{amsthm}
\usepackage{tikz}
\usetikzlibrary{arrows.meta,positioning,calc}
\usepackage[unicode,hidelinks]{hyperref}
\newcommand{\Tr}{\operatorname{Tr}}
\newcommand{\Bin}{\operatorname{Bin}}
\newcommand{\act}{\mathrm{act}}
\newcommand{\pas}{\mathrm{pas}}
\newcommand{\tagg}{\mathrm{tag}}
\newcommand{\unt}{\mathrm{untag}}
\newcommand{\ph}{\mathrm{ph}}
\newcommand{\PA}{\mathrm{PA}}
\newcommand{\eps}{\varepsilon}
\newcommand{\ket}[1]{\lvert #1\rangle}
\newcommand{\bra}[1]{\langle #1\rvert}
\newcommand{\Nsift}{N_{\mathrm{sift}}}
\newcommand{\Nunt}{N_{\mathrm{untag}}}
\newcommand{\Ntag}{N_{\mathrm{tag}}}
\newcommand{\NtagU}{\overline N_{\mathrm{tag}}}
\newcommand{\Nerr}{N_{X,\mathrm{err}}}
\newcommand{\NerrU}{N_{X,\mathrm{err}}^{\mathrm{untag}}}
\newcommand{\NerrObs}{N_{X,\mathrm{err}}^{\mathrm{obs}}}
\newcommand{\Ncross}{N_{Z,\mathrm{cross}}}
\newcommand{\Naz}{N_Z^A}
\theoremstyle{plain}
\newtheorem{theorem}{Theorem}
\newtheorem{proposition}{Proposition}
\newtheorem{mainlemma}{Lemma}
\newtheorem{lemma}[mainlemma]{Lemma}
\theoremstyle{definition}
\newtheorem{assumption}{Assumption}
\newtheorem{definition}{Definition}

\begin{document}
\title{Finite-key security of BB84 with passive receivers via reduction to active measurements}

\author{Go Kato}
\affiliation{Quantum ICT Laboratory, NICT,
4-2-1 Nukui-kitamachi, Koganei-shi, Tokyo 184-8795, Japan}

\author{Akihiro Mizutani}
\affiliation{Faculty of Engineering, University of Toyama, Gofuku 3190, Toyama 930-8555, Japan}

\author{Toyohiro Tsurumaru}
\email[Contact author: ]{Tsurumaru.Toyohiro@da.MitsubishiElectric.co.jp}
\affiliation{Mitsubishi Electric Corporation, Information Technology R\&D Center,
5-1-1 Ofuna, Kamakura-shi, Kanagawa 247-8501, Japan}

\begin{abstract}
In BB84 with passive receiver basis choice (passive BB84) and threshold detectors, the distribution of detection bases can depend on the incoming photon number. Its relation to BB84 with active receiver basis choice (active BB84), in which the basis is chosen using independent randomness, is therefore nontrivial. We establish a reduction theorem that expresses the secret-key-length formula for passive BB84 in terms of an active BB84 formula. The theorem allows a passive key-length formula to be obtained by simple substitutions into an existing active formula, without constructing a passive-specific security proof from scratch. It applies to key-length formulas derived using the Shor--Preskill, Koashi, and Renner approaches, because the reduction uses only properties shared by these approaches. Our result illustrates a modular approach that separates receiver analysis from the security proof and allows the proof of a common reference protocol to be reused for different implementations.
\end{abstract}
\maketitle

\section{Introduction}
\label{sec:introduction}
Security proofs for the Bennett--Brassard 1984 (BB84) quantum key distribution (QKD) protocol~\cite{BB84} have been developed using entanglement distillation~\cite{LoChau,ShorPreskill}, the complementarity approach~\cite{Koashi}, and the leftover hashing approach~\cite{RennerKoenig,RennerThesis,LHL,Tomamichel,TomamichelLeverrier}.
The relation between proofs based on phase error correction and leftover hashing has also been established~\cite{TsurumaruUnified}.

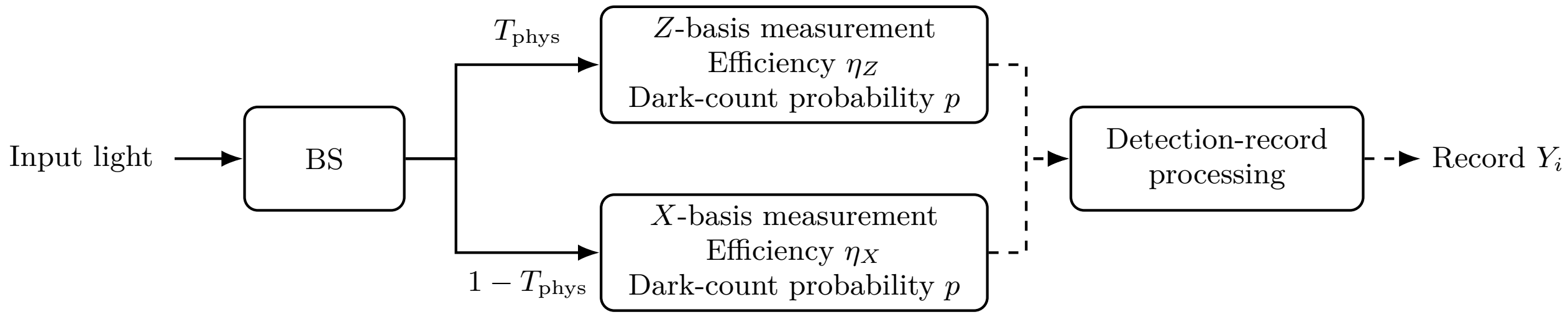
\begin{figure*}[!t]
\centering
\begin{tikzpicture}[
  >=Latex,thick,font=\small,
  box/.style={draw,rounded corners,align=center,minimum height=11mm}
]
\node[anchor=east] (in) at (-0.2,0) {Input light};
\node[box,minimum width=17mm] (bs) at (1.5,0) {BS};
\node[box,text width=39mm] (z) at (6.5,1.0)
  {$Z$-basis measurement\\Efficiency $\eta_Z$\\Dark-count probability $p$};
\node[box,text width=39mm] (x) at (6.5,-1.0)
  {$X$-basis measurement\\Efficiency $\eta_X$\\Dark-count probability $p$};
\node[box,text width=29mm] (rec) at (11.0,0) {Detection-record processing};
\draw[->,shorten <=3pt] (in.east)--(bs.west);
\draw[->] (bs.east)--(2.9,0)--(2.9,1.0)
  --node[midway,above=2pt] {$T_{\rm phys}$} (z.west);
\draw[->] (bs.east)--(2.9,0)--(2.9,-1.0)
  --node[midway,below=2pt] {$1-T_{\rm phys}$} (x.west);
\draw[->,dashed] (z.east)--++(0.4,0)|-(rec.west);
\draw[->,dashed] (x.east)--++(0.4,0)|-(rec.west);
\draw[->,dashed] (rec.east)--++(0.6,0) node[right] {Record $Y_i$};
\end{tikzpicture}
\caption{Actual passive receiver. Each basis path contains two threshold detectors. Detection-record processing is specified in Sec.~\ref{sec:protocol} and produces the measurement record $Y_i$ in Eq.~\eqref{eq:record}. In Figs.~\ref{fig:physical}--\ref{fig:F}, solid and dashed lines indicate optical systems and classical information, respectively.}
\label{fig:physical}
\end{figure*}

Many existing finite-key formulas assume BB84 with active receiver basis choice (active BB84), in which Bob chooses his measurement basis using independent randomness~\cite{Tomamichel,HayashiTsurumaru,TomamichelLeverrier}.
In BB84 with passive receiver basis choice (passive BB84), by contrast, Bob's receiver splits the incoming light at a beam splitter (BS) and determines the basis from the detection record (Fig. \ref{fig:physical}).
Passive basis choice eliminates the need to generate random numbers for Bob's basis selection, to drive a basis switch at high speed, and to synchronize its drive signal with the incoming pulses. 
Because of these practical advantages, passive basis choice is used in various implementations, including terrestrial QKD and quantum networks~\cite{Morrison2023,Gruenenfelder2023,Lu2025} and satellite QKD~\cite{LiSatellite2025}.

In security proofs for active BB84,
the detectors are typically assumed to have no dark counts.
This is because, for the detector model considered here,
dark counts can be simulated by a basis-independent quantum channel
and interpreted as part of Eve's attack (see Sec.~\ref{sec:dark}).
However, this construction does not directly extend to passive BB84.
For example, when Bob's detectors have unit efficiency and no dark counts, an $n$-photon input produces detections exclusively in the $Z$ or $X$ path with probabilities $T^n$ and $(1-T)^n$, respectively, where $T$ is the BS transmittance toward the $Z$ path.
Conditioned on detection in only one path, the basis distribution generally depends on $n$.
Even if Alice sends a single photon, Eve can send a multiphoton state to Bob.
Thus, simply replacing the BS by independent basis selection with probabilities $T$ and $1-T$ does not reproduce the detection statistics.

Existing security analyses for passive receivers include flag-state squashing methods~\cite{Kamin}, analytical asymptotic proofs~\cite{Kawakami}, and finite-key analyses~\cite{Mizutani,WangPassive}.
For example, Ref.~\cite{Mizutani} uses the complementarity approach to relate phase errors to cross clicks, which are events in which both basis paths register detections.
References~\cite{WangPassive,Nahar2026} present security analyses of three-state time-bin QKD with passive basis choice in combination with flag-state squashing.

In this paper, we establish a theorem that reduces the secret-key-length formula for passive BB84 to an active BB84 formula (Theorem~\ref{thm:transfer}).
This avoids the need to construct a passive-specific security proof from scratch: a passive key-length formula is obtained by simple substitutions into an existing active formula.

The reduction applies across widely used security-proof approaches.
In particular, active BB84 key-length formulas derived using the Shor--Preskill approach~\cite{ShorPreskill}, Koashi's approach~\cite{Koashi,HayashiTsurumaru}, or Renner's approach~\cite{RennerThesis,Tomamichel} can be converted into passive BB84 formulas.
The proof of the reduction uses only properties shared by these approaches, collected in Assumption~\ref{ass:active-formula}.

The proof of Theorem~\ref{thm:transfer} proceeds as follows. 
First, we show that the theoretical model of the passive BB84 receiver is equivalent to a quantum operation $F$ followed by an active BB84 receiver.
This makes passive BB84 equivalent to an active protocol with an additional operation that inserts tagged bits, treated as known to Eve, into the sifted key [Fig.~\ref{fig:tag-insertion}(b)].
Once an upper bound $\NtagU$ on the number of tagged bits is obtained, suitable substitutions into the active BB84 formula give a passive BB84 key-length formula.

The bound $\NtagU$ can be calculated by classical interval estimation from the observed number of cross clicks.
The substitution into the active formula is also common to the different proof approaches.
For the Shor--Preskill and Koashi approaches, it follows naturally from the idea of tagged qubits~\cite{GLLP}.
For Renner's approach, the same substitution is justified by properties of conditional min-entropy and the leftover hashing lemma.
The resulting reduction theorem therefore applies to all three approaches.

This paper is organized as follows.
Section~\ref{sec:setting} specifies the devices, protocol, and security criterion, and Sec.~\ref{sec:main-result} presents the reduction theorem.
Section~\ref{sec:idea} constructs the receiver decomposition, and Sec.~\ref{sec:proof} proves the reduction theorem.
The appendices give the security arguments for tag insertion, a binomial method for bounding the tag count and its closed-form relaxation, and an explicit finite-key formula.

\section{Setting and protocol}
\label{sec:setting}
We first specify the passive BB84 protocol considered in this paper.

\subsection{Source and receiver}
\label{sec:receiver}

The transmitter is the same as in active BB84.
We use two specified optical modes and write $\ket{n_0,n_1}$ for the state
with $n_0$ and $n_1$ photons in the respective modes.
The $Z$ basis is defined by $\ket{0_Z}=\ket{1,0}$ and $\ket{1_Z}=\ket{0,1}$,
and the $X$ basis by
$\ket{b_X}=(\ket{0_Z}+(-1)^b\ket{1_Z})/\sqrt2$,
where $b\in\{0,1\}$.
The $Z$ basis is used for key generation and the $X$ basis for phase-error estimation.
In each round, Alice independently chooses a basis $\alpha_i\in\{Z,X\}$
and a uniformly random bit $a_i\in\{0,1\}$,
and sends the corresponding ideal single-photon state.
The probability of choosing the $Z$ basis is $p_Z^A\in(0,1)$.

The receiver is specified as follows.
\begin{definition}[Passive BB84 receiver, Fig.~\ref{fig:physical}]
\label{def:receiver}
A passive BB84 receiver satisfies the following conditions.
\begin{itemize}
    \item (State space) The receiver input is an arbitrary state on the Fock space of the two optical modes defined above.
    \item (Beam splitter)
    A BS with vacuum at its auxiliary input routes each incoming photon to the $Z$ and $X$ paths with probabilities $T_{\rm phys}$ and $1-T_{\rm phys}$, respectively.
    The splitting ratio is independent of the allowed input mode.
    \item (Detectors) Each of the $Z$ and $X$ paths consists of a linear optical transformation for the corresponding BB84 basis and two threshold detectors associated with bits $0$ and $1$.
    \begin{itemize}
    \item (Efficiencies) The two detectors within each path have equal detection efficiencies.
    
    This condition allows us to define the overall efficiencies of the two measurement paths (including the detector quantum efficiencies, optical losses in the respective measurement paths, and losses due to time-window selection~\cite{NoteTimeBinEfficiency}, etc.), which we denote by $\eta_Z,\eta_X\in(0,1]$.
    \item (Dark counts) Each detector produces a dark count with probability $p\in[0,1)$, independently of photon detection.
    Dark counts are independent across detectors and transmission rounds.
    \end{itemize}
    \item (Effective BS transmittance) The effective beam splitter (BS) transmittance $T$, defined by
    \begin{align}
    T&=\frac{T_{\rm phys}\eta_Z}{\eta_{\rm in}},
    \label{eq:effective-T}\\
    \eta_{\rm in}
    &=T_{\rm phys}\eta_Z+(1-T_{\rm phys})\eta_X,
    \label{eq:eta_in}
    \end{align}
    satisfies $1/2\le T<1$.
    The physical meaning of $T$ is explained in the paragraph below.
    \item (Detection records) If detections occur in only one path, the basis and bit for that path are recorded.

    A double click, in which both detectors in the same path respond, is assigned a uniformly random bit.

    A cross click, in which detections occur in both paths, is recorded by its round index and is used neither for the sifted key nor for the sample.
\end{itemize}
\end{definition}
The effective BS transmittance $T$ was defined as in (\ref{eq:effective-T}) due to the following observation.
By the properties of linear optics, the effects of transmittances $T_{\rm phys},\eta_Z,\eta_X$ can be rewritten as the common input loss $\eta_{\rm in}$ of (\ref{eq:eta_in}) followed by BS transmittance $T$.
This makes the receiver in Definition~\ref{def:receiver} equivalent to a model consisting
of an input loss with transmittance $\eta_{\rm in}$,
followed by a BS with transmittance $T$ and unit-efficiency detectors~\cite{NoteEffectiveEfficiency}. 

\subsection{Actual passive BB84 protocol}
\label{sec:protocol}

The protocol follows the same outline as active BB84:
Alice sends BB84 states and Bob receives them.
The sifted keys $K_A,K_B$ are their respective bit strings from rounds in which
Alice chooses $Z$ and Bob registers detections only in the $Z$ path.
Similarly, their bit strings from rounds in which Alice chooses $X$ and Bob
registers detections only in the $X$ path form the sample.

Bob's passive basis choice leads to the following difference from the active protocol.
His measurement record in round $i$ is
\begin{equation}
 Y_i\in\{Z0,Z1,X0,X1,\mathrm{cross},\emptyset\}.
 \label{eq:record}
\end{equation}
Here, $Zb$ ($Xb$) denotes a detection in the $Z$ ($X$) path only,
with recorded bit $b\in\{0,1\}$.
For a single click in that path, the bit associated with the detector that responds is recorded.
For a double click, a uniformly random bit is recorded.
The outcome $\mathrm{cross}$ denotes detections in both paths,
and $\emptyset$ denotes no response from any detector.

The detailed protocol is as follows. All classical communication is authenticated.
\begin{enumerate}
\item \textbf{Preparation and quantum communication.}
Before quantum communication, fix the number of transmitted signals $N$,
the probability $p_Z^A$, the receiver settings, and the security parameters.

Also fix the statistical estimation method used for the key-length calculation
in Sec.~\ref{sec:main}.

Alice sends $N$ signals, and Bob performs the measurement in Definition~\ref{def:receiver}.
No bases, detection outcomes, or bit values are announced until all $N$ rounds
of reception have been completed.

\item \textbf{Information exchange.}
After all $N$ transmission rounds have been completed,
Alice announces her bases and Bob announces his detection bases
and the positions of cross clicks and no-detection outcomes. 
They reveal their $X$-sample bit values and count the sample errors $\NerrObs$.
The sifted-key length $\Nsift$ is the number of rounds in which Alice chose $Z$
and Bob registered detections only in the $Z$ path.
We also write $\Naz$ for the number of rounds in which Alice chose $Z$,
irrespective of Bob's detection outcome, and $\Ncross$ for the number of cross clicks
among those rounds.

\item \textbf{Error correction and error verification.}
Apply the error correction and error verification specified by the chosen
active BB84 security proof to the sifted keys $K_A,K_B$.
Abort the protocol if error verification fails.
Let $\lambda_{\rm EC}$ be an upper bound on the public information leakage
from error correction and error verification.
Use the leakage evaluation specified by the original active proof,
including its treatment of the public verification hash and the pass/fail outcome.

\item \textbf{Privacy amplification (PA).}
Calculate the output length $\ell$ using the formulas in Sec.~\ref{sec:main}.
Abort if $\ell\le0$; otherwise, perform the PA specified by the original active BB84 protocol.
Choose the PA randomness independently after determining the output length.
Denote Alice's and Bob's final keys by $S_A,S_B$ and set
$S_A=S_B=\bot$ upon abort.
\end{enumerate}

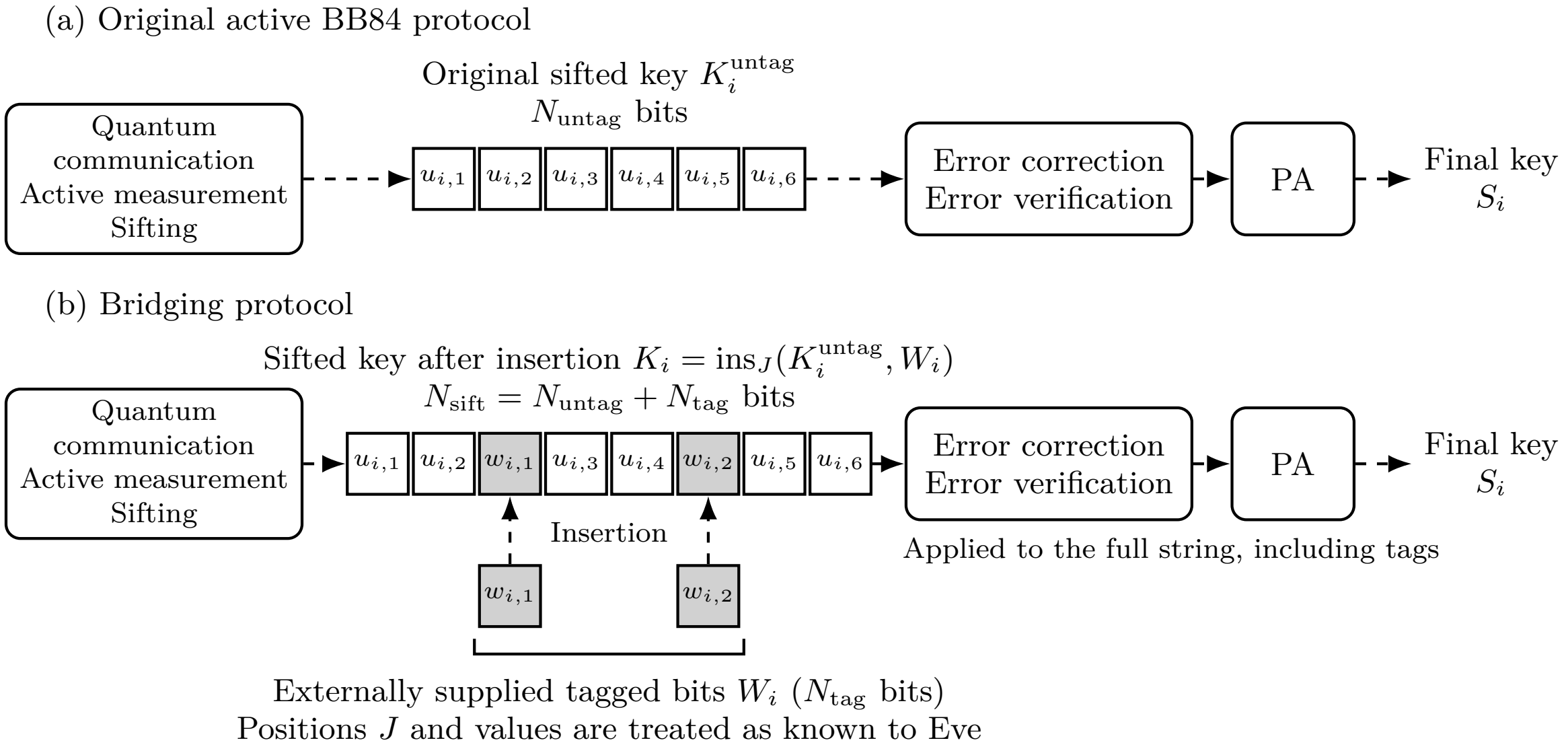
\begin{figure*}[!t]\centering
\begin{tikzpicture}[>=Latex,thick,font=\small,
 box/.style={draw,rounded corners,align=center,minimum height=11mm},
 bit/.style={draw,minimum width=6mm,minimum height=6mm,inner sep=0pt,font=\scriptsize},
 tagbit/.style={bit,fill=black!18}]

\node[anchor=west] at (0,1.55) {(a) Original active BB84 protocol};
\node[box,text width=27mm,font=\footnotesize] (quantum-a) at (1.2,0)
 {Quantum\\communication\\Active measurement\\Sifting};
\node[align=center] at (5.675,.85)
 {Original sifted key $K_i^{\unt}$\\$\Nunt$ bits};

\foreach \k/\x in {1/4.05,2/4.70,3/5.35,4/6.00,5/6.65,6/7.30}{
 \node[bit] (a\k) at (\x,0) {$u_{i,\k}$};
}

\node[box,text width=26mm] (ec-a) at (10.0,0) {Error correction\\Error verification};
\node[box,minimum width=12mm] (pa-a) at (12.4,0) {PA};
\node[align=center] (out-a) at (14.35,0) {Final key\\$S_i$};

\draw[->,dashed] (quantum-a.east)--(a1.west);
\draw[->,dashed] (a6.east)--(ec-a.west);
\draw[->,dashed] (ec-a.east)--(pa-a.west);
\draw[->,dashed] (pa-a.east)--(out-a.west);

\node[anchor=west] at (0,-1.25) {(b) Bridging protocol};
\node[box,text width=27mm,font=\footnotesize] (quantum-b) at (1.2,-2.80)
 {Quantum\\communication\\Active measurement\\Sifting};
\node[align=center] at (5.675,-1.95)
 {Sifted key after insertion $K_i=\operatorname{ins}_J(K_i^{\unt},W_i)$\\
 $\Nsift=\Nunt+\Ntag$ bits};

\foreach \k/\x in {1/3.40,2/4.05,3/5.35,4/6.00,5/7.30,6/7.95}{
 \node[bit] (b\k) at (\x,-2.80) {$u_{i,\k}$};
}

\node[tagbit] (t1) at (4.70,-2.80) {$w_{i,1}$};
\node[tagbit] (t2) at (6.65,-2.80) {$w_{i,2}$};
\node[box,text width=26mm] (ec-b) at (10.0,-2.80) {Error correction\\Error verification};
\node[box,minimum width=12mm] (pa-b) at (12.4,-2.80) {PA};
\node[align=center] (out-b) at (14.35,-2.80) {Final key\\$S_i$};

\draw[->,dashed] (quantum-b.east)--(b1.west);
\draw[->,dashed] (b6.east)--(ec-b.west);
\draw[->,dashed] (ec-b.east)--(pa-b.west);
\draw[->,dashed] (pa-b.east)--(out-b.west);

\node[font=\footnotesize,align=center] at (11.2,-3.65)
 {Applied to the full string, including tags};

\node[tagbit] (w1) at (4.70,-4.10) {$w_{i,1}$};
\node[tagbit] (w2) at (6.65,-4.10) {$w_{i,2}$};

\draw[->,dashed] (w1.north)--(t1.south);
\draw[->,dashed] (w2.north)--(t2.south);

\node[font=\footnotesize] at (5.675,-3.47) {Insertion};
\draw (4.35,-4.52)--(4.35,-4.67)--(7.00,-4.67)--(7.00,-4.52);
\node[align=center,anchor=north] at (5.675,-4.78)
 {Externally supplied tagged bits $W_i$ ($\Ntag$ bits)\\
 Positions $J$ and values are treated as known to Eve};

\end{tikzpicture}
\caption{(a) The original active BB84 protocol (Sec.~\ref{sec:original_BB84_description}) and (b) the bridging protocol (Sec.~\ref{sec:bridging_protocol_description}).
In the bridging protocol, tags (also called tagged bits) are inserted into the sifted key obtained from quantum communication before key distillation.
The index $i=A,B$ denotes Alice or Bob.
The white boxes $u_{i,k}$ represent the $k$th bits of the original sifted key $K_i^{\unt}$, and the gray boxes $w_{i,k}$ represent the $k$th bits of the inserted string $W_i$ [Eq.~\eqref{eq:insertion}].
The example has $\Nunt=6$, $\Ntag=2$, and $J=\{3,6\}$; insertion precedes error correction.
Alice and Bob use the same insertion positions, but the corresponding bit values need not agree.
No secrecy is required of the tags. 
Key distillation (error correction, error verification, and PA) acts on the entire $\Nsift$-bit sifted key, including the tags.}
\label{fig:tag-insertion}
\end{figure*}

\begin{figure*}[!t]\centering
\begin{tikzpicture}[>=Latex,thick,font=\small,box/.style={draw,rounded corners,align=center,minimum height=16mm}]
\node (in) at (0,0) {Input light};
\node[box,minimum width=13mm] (f) at (2.0,0) {$F$};
\node[box,text width=33mm] (act) at (5.6,0) {Ideal active measurement\\No dark counts\\Unit efficiency\\$Z:X=T:(1-T)$};
\node[box,text width=35mm] (rec) at (10.8,0) {Record selection\\untag: measurement\\outcome\\tag: basis and bit from $c$\\cross: cross};
\draw[->] (in)--(f);
\draw[->] (f.east)--node[midway,above,align=center] {Optical\\system} (act.west);
\draw[->,dashed] (act)--node[above] {Outcome} (rec);
\draw[->,dashed] (f.south)--++(0,-1.25)-|node[pos=.3,below] {Classical flag $c$} (rec.south);
\draw[->,dashed] (rec.east)--++(.7,0) node[right] {$Y$};
\end{tikzpicture}
\caption{
Virtual receiver equivalent to the actual passive receiver
of Definition~\ref{def:receiver}.
The record $Y$ is obtained from the ideal
active measurement on an untagged branch and from the classical
flag $c$ on a tagged or cross branch.
For every input state, the two receivers produce the same
joint state of $Y$ and any systems correlated with the input,
as established by Lemma~\ref{lem:exact}.
}
\label{fig:modified}
\end{figure*}
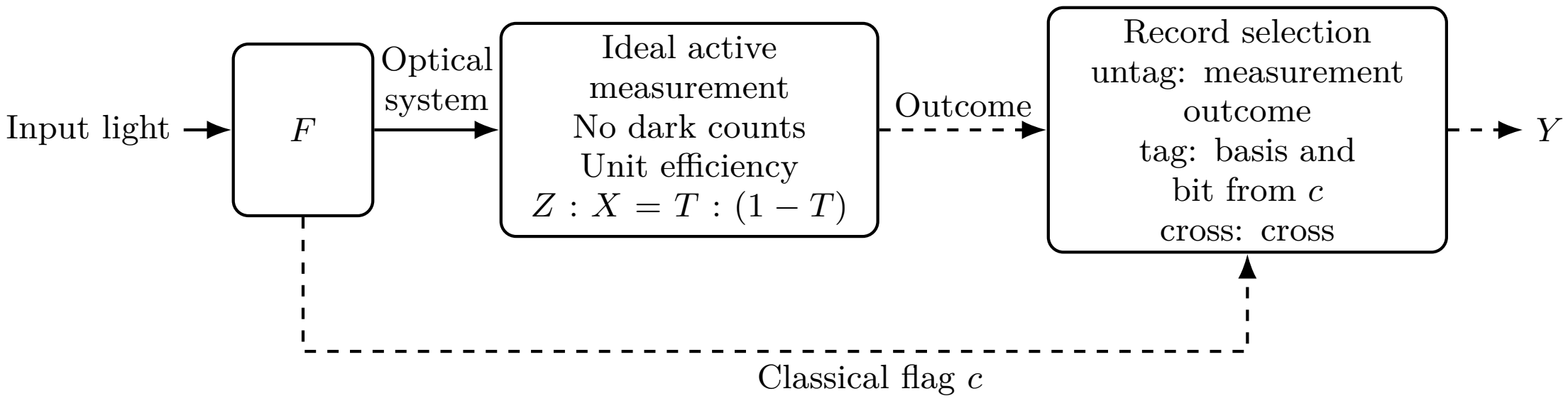

\subsection{Security definition}
\label{sec:security-definition}

Let $C$ denote the complete public transcript of the actual protocol defined in
Sec.~\ref{sec:protocol}, and let $E$ denote Eve's system.
The final state is denoted by $\rho_{S_AS_BEC}$.
The transcript $C$ includes the final key length and the pass/fail outcome.
Define $\rho^{\rm ideal}_{S_AS_BEC}$ by replacing the keys in successful executions
with identical, uniformly random keys of the same length that are independent of $EC$,
while preserving the public transcript, abort outcomes, and Eve's marginal state.

\begin{definition}[QKD security]
\label{def:security}
A protocol is $\eps$-secure if
\begin{equation}
 d(\rho,\rho^{\rm ideal})
 :=\tfrac12\|\rho_{S_AS_BEC}-\rho_{S_AS_BEC}^{\rm ideal}\|_1\le\eps
 \label{eq:qkd-security}
\end{equation}
for every attack by Eve. 
Here, $\|A\|_1:=\Tr\sqrt{A^\dagger A}$ denotes the trace norm,
and $d$ is the trace distance. 
\end{definition}

This is composable security for the entire execution, including variable key lengths
and aborts \cite{PortmannRenner,TomamichelLeverrier}.
Explicit expressions for the real and ideal states are given in Appendix~\ref{app:security}.

\section{Main result: Reduction from active to passive BB84}\label{sec:main-result}
In this section, we present our main result.
It shows how to obtain a secret-key-length formula for passive BB84
by making simple substitutions in an existing formula for active BB84.
The argument has three steps.
\begin{enumerate}
    \item \textbf{Define a bridging protocol.}
    To reduce the security of passive BB84 to that of active BB84, we extend the active protocol by inserting tagged bits into the sifted key obtained from quantum communication, before key distillation (Fig.~\ref{fig:tag-insertion}).
    We call this extension the bridging protocol.
    Tagged bits are additional bits whose values and insertion positions are treated as known to Eve.

    \item \textbf{Obtain the key length of the bridging protocol from the original protocol.}
    For the widely used proof approaches in Refs.~\cite{ShorPreskill,Koashi,RennerThesis,Tomamichel,HayashiTsurumaru}, with universal$_2$ hashing for privacy amplification, the key-length formula for the bridging protocol follows directly from the original formula (Sec.~\ref{sec:formula}).

    \item \textbf{Obtain the key length of passive BB84 from the bridging protocol.}
    The passive receiver can be decomposed into a quantum operation $F$ followed by ideal active measurements (Fig.~\ref{fig:modified}).
    This decomposition represents passive BB84 as a bridging protocol.
    Applying step 2 then gives the secret-key length of passive BB84 (Sec.~\ref{sec:main}).
\end{enumerate}

Sections~\ref{sec:active-input} and \ref{sec:formula} describe the bridging protocol and its key length, respectively.
Section~\ref{sec:main} states the main theorem.

\subsection{Bridging protocol}\label{sec:active-input}
We define the bridging protocol as a modification of active BB84.

\subsubsection{Original active BB84 protocol}\label{sec:original_BB84_description}
We start from the active BB84 protocol in Fig.~\ref{fig:tag-insertion}(a).
Alice sends randomly chosen BB84 states, and Bob measures each signal in the $Z$ or $X$ basis, chosen with probabilities $p_Z^B$ and $1-p_Z^B$, respectively.
A uniformly random bit is assigned to each double click.
The detectors have zero dark counts and unit efficiency~\cite{NoteActiveUnitEfficiency}.
We write Alice's and Bob's basis-choice probabilities as
\begin{equation}
\boldsymbol p_Z:=(p_Z^A,p_Z^B).
\end{equation}

\subsubsection{Bridging protocol}\label{sec:bridging_protocol_description}
The bridging protocol adds a virtual operation that inserts externally supplied bit strings into the sifted keys and sample strings 
of the original active protocol [Fig.~\ref{fig:tag-insertion}(b)].

\paragraph{Operation on the sifted keys.}
We first describe the insertion into the sifted keys.
\begin{itemize}
    \item The sifted keys obtained from quantum communication in the original protocol are called the untagged sifted keys and denoted by $K_A^{\unt},K_B^{\unt}$, with $|K_A^{\unt}|=|K_B^{\unt}|=:\Nunt$.
    \item Eve generates the tagged bit strings $W_A,W_B$, with $|W_A|=|W_B|=:\Ntag$.
    \item Alice and Bob insert $W_A,W_B$ into $K_A^{\unt},K_B^{\unt}$, respectively, and use the resulting strings as their sifted keys $K_A,K_B$.
    Writing $|K_A|=|K_B|=\Nsift$ gives
    \begin{equation}
    \Nsift=\Nunt+\Ntag.
    \label{eq:total_sifted_key_length}
    \end{equation}
\end{itemize}
We also refer to $K_A^{\unt},K_B^{\unt}$ as the original keys and to $W_A,W_B$ as the tags.
Before the basis announcement, Eve specifies the transmission round,
basis, and Alice's and Bob's bit values for each tag candidate.
The candidate specifies Bob's recorded basis,
while Alice uses her preparation basis chosen in the original protocol.
After the announcement, Alice and Bob retain candidates for which
Alice's preparation basis and Bob's recorded basis are both $Z$.
The retained bit values, arranged in transmission order,
form $W_A$ and $W_B$. 
The tag positions in the resulting sifted keys are represented by
$J\subseteq\{1,\ldots,\Nsift\}$, with $|J|=\Ntag$.
The resulting keys are
\begin{equation}
 K_A=\operatorname{ins}_J(K_A^{\unt},W_A),\quad
 K_B=\operatorname{ins}_J(K_B^{\unt},W_B),
 \label{eq:insertion}
\end{equation}
where $\operatorname{ins}_J$ preserves the order of all original bits and places the bits of $W_A,W_B$ in order at the positions in $J$.

\paragraph{Operation on the sample strings.}
Tagged data are similarly added to the sample strings, namely, the $X$-basis measurement outcomes.
The error count $\NerrU$ in the untagged part and the total observed error count $\NerrObs$ satisfy
\begin{equation}
 \NerrObs=\NerrU+N_{X,\mathrm{err}}^{\tagg}\ge\NerrU.
 \label{eq:sample-addition}
\end{equation}

\paragraph{Key distillation.}
After insertion, Alice and Bob apply error correction, error verification,
and PA to $K_A,K_B$ in Eq.~\eqref{eq:insertion} to generate their secret keys.

\subsection{Assumptions on the original active BB84 key-length formula}
\label{sec:formula}
We require the secret-key-length formula for active BB84 to satisfy Assumption~\ref{ass:active-formula} below.
When PA uses universal$_2$ hashing, the familiar proof approaches of Shor--Preskill~\cite{ShorPreskill}, Koashi~\cite{Koashi,HayashiTsurumaru}, and Renner~\cite{RennerThesis,Tomamichel} satisfy this assumption, as shown in Appendix~\ref{sec:well-known-methods}.

\begin{assumption}[Active BB84 key-length formula]\label{ass:active-formula}
The secret-key-length formula for active BB84 satisfies the following two conditions.
\begin{enumerate}
\item (Original active BB84 protocol)
The active BB84 protocol in Sec.~\ref{sec:original_BB84_description} is $\eps$-secure when its secret-key length satisfies
\begin{align}
\ell_\act&\le\mathcal L_\act\big(N,\boldsymbol p_Z,\Nsift,\Nerr,\lambda_{\rm EC};\eps\big)\nonumber\\
&:=\Nsift-\Gamma\big(N,\boldsymbol p_Z,\Nsift,\Nerr;\eps\big)-\lambda_{\rm EC}.
\label{eq:normalBB84-KL}
\end{align}
Here $N$ is the fixed number of transmitted signals specified in Sec.~\ref{sec:protocol}, $\Nsift$ is the sifted-key length, and $\Nerr$ is the error count in the $X$ sample.
With all other arguments fixed,
\[
 \Nsift-\Gamma\bigl(N,\boldsymbol p_Z,\Nsift,\Nerr;\eps\bigr)
\]
is nondecreasing in the sifted-key length $\Nsift$ and nonincreasing in the sample error count $\Nerr$.

\item (Reuse for the bridging protocol)
The bridging protocol is $\eps$-secure when its secret-key length satisfies
\begin{align}
\ell_{\rm bri}&\le\mathcal L_{\rm bri}\big(N,\boldsymbol p_Z,\Nsift,\Ntag,\NerrObs,\lambda_{\rm EC};\eps\big)\nonumber\\
&:=\mathcal L_\act\big(N,\boldsymbol p_Z,\Nunt,\NerrObs,\lambda_{\rm EC};\eps\big).
\label{eq:normal-bridge-BB84KL_relation}
\end{align}
The sifted-key length $\Nsift$ and sample error count $\NerrObs$ in Eq.~\eqref{eq:normal-bridge-BB84KL_relation} include tagged bits [Eqs.~\eqref{eq:total_sifted_key_length} and \eqref{eq:sample-addition}].

Since $\Nunt=\Nsift-\Ntag$ in Eq.~\eqref{eq:normal-bridge-BB84KL_relation}, the monotonicity in condition 1 implies that $\mathcal L_{\rm bri}$ is nonincreasing in $\Ntag$ when all other arguments, including the total sifted-key length $\Nsift$, are fixed.
Thus, overestimating the number of tags cannot increase the key-length bound.
\end{enumerate}
\end{assumption}

The term $\Gamma$ in Eq.~\eqref{eq:normalBB84-KL} accounts for the information available to the adversary and specifies the number of bits removed by PA.
It is sometimes called the sacrifice-bit length~\cite{HayashiTsurumaru}.
Appendix~\ref{app:example-formula} gives a concrete example of $\Gamma$.

To interpret condition 2, apply $\NerrObs\ge\NerrU$ from Eq.~\eqref{eq:sample-addition} and the nonincreasing dependence on the sample error count in condition 1 to Eqs.~\eqref{eq:normalBB84-KL} and \eqref{eq:normal-bridge-BB84KL_relation}.
The bridging formula then satisfies
\begin{align}
&\mathcal L_{\rm bri}\big(N,\boldsymbol p_Z,\Nsift,\Ntag,\NerrObs,\lambda_{\rm EC};\eps\big)\nonumber\\
&=\Nunt-\Gamma\big(N,\boldsymbol p_Z,\Nunt,\NerrObs;\eps\big)-\lambda_{\rm EC}\nonumber\\
&\le\mathcal L_\act \big(N,\boldsymbol p_Z,\Nunt,\NerrU,\lambda_{\rm EC};\eps\big).
\label{eq:bridge-BB84KL}
\end{align}
The last line of Eq.~\eqref{eq:bridge-BB84KL} evaluates the original
active formula using the untagged key length $\Nunt$ and the
untagged sample error count $\NerrU$.
Since $\Nunt=\Nsift-\Ntag$, this expression corresponds to applying
PA to the entire sifted key of length $\Nsift$, with $\Ntag$ added
to the PA sacrifice required for the untagged part.
On both sides of Eq.~\eqref{eq:bridge-BB84KL}, $\lambda_{\rm EC}$
bounds the public leakage from error correction and error
verification on the entire sifted key, including tags.

\begin{table*}[!t]
\centering
\caption{
Substitutions for calculating the key length from observations in the actual passive protocol. 
The substitutions into the bridging formula and the original active formula correspond to Eqs.~\eqref{eq:key-transfer2} and \eqref{eq:key-transfer}, respectively.
}
\label{tab:substitution}
\begin{tabular}{@{}p{.22\textwidth}p{.28\textwidth}p{.44\textwidth}@{}}
\toprule
Formula & Input & Value for the actual passive protocol\\
\midrule
Both formulas
& Number of transmitted signals
& $N$\\
Both formulas
& Basis-choice probabilities $\boldsymbol p_Z$
& $(p_Z^A,T)$\\
Bridging formula $\mathcal L_{\rm bri}$
& Total sifted-key length
& $\Nsift$\\
Bridging formula $\mathcal L_{\rm bri}$
& Number of tags
& $\NtagU$\\
Original active formula $\mathcal L_{\act}$
& Sifted-key length (untagged part)
& $\Nsift-\NtagU$\\
Both formulas
& Sample error count
& $\NerrObs$\\
Both formulas
& Public leakage $\lambda_{\rm EC}$
& Leakage bound for error correction and error verification on the entire sifted key, including tags\\
Both formulas
& Security parameter
& The same predetermined value of $\eps$\\
\bottomrule
\end{tabular}
\end{table*}

\subsection{Main theorem and substitution of observed quantities}\label{sec:main}
As proved in the following sections, passive BB84 can be represented as a bridging protocol. 
However, Alice and Bob cannot directly observe the number $\Ntag$ of tagged sifted-key bits in the actual passive implementation. 
They therefore need an upper bound $\NtagU$ with failure probability at most $\delta_{\tagg}$, satisfying
\begin{equation}
 0\le\NtagU\le\Nsift,\quad
 \Pr[\Ntag>\NtagU]\le\delta_{\tagg}.
 \label{eq:coverage}
\end{equation}
Given an active BB84 secret-key-length formula $\mathcal L_\act$ and such an upper bound $\NtagU$, we obtain a formula $\mathcal L_{\rm pas}$ for passive BB84.

\begin{theorem}[Transfer of a key-length formula from active to passive BB84]\label{thm:transfer}
Suppose that an active BB84 secret-key-length formula $\mathcal L_\act$ satisfies Assumption~\ref{ass:active-formula} and that $\NtagU$ upper-bounds $\Ntag$ except with probability at most $\delta_{\tagg}$.
Then passive BB84 is $(\eps+\delta_{\tagg})$-secure when its secret-key length satisfies
\begin{align}
\ell_{\pas}&\le
\mathcal L_{\rm pas}\big(N,p_Z^A,T,\Nsift,\NtagU,\NerrObs,\lambda_{\rm EC};\eps\big)\nonumber\\
&:=\mathcal L_{\rm bri}\big(N,(p_Z^A,T),\Nsift,\NtagU,\NerrObs,\lambda_{\rm EC};\eps\big)
\label{eq:key-transfer2}
\\
&=\mathcal L_\act\big(N,(p_Z^A,T),\Nsift-\NtagU,\NerrObs,\lambda_{\rm EC};\eps\big).
\label{eq:key-transfer}
\end{align}
\end{theorem}
Equation~\eqref{eq:key-transfer2} gives the passive formula by substituting the parameters of passive BB84 into the bridging formula, as summarized in Table~\ref{tab:substitution}. 
Using Eq.~\eqref{eq:normal-bridge-BB84KL_relation}
in Eq.~\eqref{eq:key-transfer2} gives Eq.~\eqref{eq:key-transfer},
which expresses the passive formula directly in terms of
the original active formula.

Appendix~\ref{app:statistics} gives a general procedure for calculating $\NtagU$.
As an example, we give below a closed-form upper bound $\NtagU^{\rm exmp}$ that requires no numerical search.
For $1/2<T<1$ and $0<\delta_{\tagg}<1$, set
$r:=2(1-T)$ and $s_{\tagg}:=\ln(1/\delta_{\tagg})$.
Using the observed counts $\Ncross,\Nsift$ defined in Sec.~\ref{sec:protocol}, let
\begin{equation}
\begin{split}
\NtagU^{\rm exmp}
:=\min\Biggl\{\Nsift,\,
\Biggl\lfloor
&\frac{(1-r)\Ncross+s_{\tagg}}{r}\\
&+\frac{
\sqrt{s_{\tagg}^{\,2}+2s_{\tagg}\Ncross}
}{r}
\Biggr\rfloor
\Biggr\}.
\end{split}
\label{eq:tag-bound-closed}
\end{equation}
For $T=1/2$, set $\NtagU=0$.
This bound satisfies Eq.~\eqref{eq:coverage}.
Appendix~\ref{app:statistics} gives a procedure based directly on the binomial cumulative probability and proves Eq.~\eqref{eq:tag-bound-closed}.

\section{Receiver reduction: Main lemma and construction}
\label{sec:idea}
This section proves Lemma~\ref{lem:exact}, which establishes
the equivalence between the actual passive receiver of
Definition~\ref{def:receiver}~\cite{NoteReceiverInputLoss}
and the virtual receiver in Fig.~\ref{fig:modified}.
In the virtual receiver, a completely positive, trace-preserving
(CPTP) map $F$ is applied before basis selection and outputs
both an optical system for the ideal active measurement
and a classical flag.
The record $Y$ is determined from
the ideal measurement outcome and this flag.

We first show how to simulate dark counts by a map on the optical input (Proposition \ref{prop:dark_count_simulatability}),
and then use this result to construct $F$ so that the virtual receiver
reproduces the actual passive receiver's measurement statistics,
including correlations with other systems.
This equivalence is used to prove Theorem~\ref{thm:transfer}
in the next section.

\subsection{Simulating dark counts by a map on the input}
\label{sec:dark}
\begin{figure*}[!t]\centering
\begin{tikzpicture}[>=Latex,thick,font=\small,
 box/.style={draw,rounded corners,align=center,minimum height=8mm},
 lab/.style={anchor=west,font=\small}]

\node[box,text width=24mm] (qnd) at (0,0)
 {QND measurement\\of total photon\\number $n$};
\node[anchor=east] (bin) at (-2.4,0) {$B_{\rm in}$};
\draw[->] (bin.east)--(qnd.west);

\node[box] (zero) at (3.2,2.15) {$n=0$};
\node[box] (nonzero) at (3.2,-2.15) {$n\ge1$};
\draw[->] (qnd.east)--++(.4,0)|-(zero.west);
\draw[->] (qnd.east)--++(.4,0)|-(nonzero.west);

\node[box,text width=23mm] (spi0) at (6.8,3.35)
 {$\mathrm{SPI}_{r_0}$};
\node[box,text width=23mm] (tagx) at (6.8,2.15)
 {Generate a\\uniform bit $b$};
\node[box,text width=23mm] (cross0) at (6.8,.95)
 {Discard input};

\draw[->] (zero.east)--++(.3,0)
 |-node[pos=.74,above] {$u_0$} (spi0.west);
\draw[->] (zero.east)
 --node[midway,above=2pt] {$t_{X,0}$} (tagx.west);
\draw[->] (zero.east)--++(.3,0)
 |-node[pos=.74,below] {$c_0$} (cross0.west);

\node[lab] (o0) at (8.9,3.55)
 {Light: $\mathrm{SPI}_{r_0}(\rho_0)$};
\node[lab] (l0) at (8.9,3.15) {$c=\mathrm{untag}$};
\node[lab] (ox) at (8.9,2.35) {Light: $\rho_0$};
\node[lab] (lx) at (8.9,1.95) {$c=\mathrm{tagged}_{X,b}$};
\node[lab] (oc0) at (8.9,1.15) {Light: $\rho_0$};
\node[lab] (lc0) at (8.9,.75) {$c=\mathrm{cross}$};

\foreach \a/\b/\c in {spi0/o0/l0,tagx/ox/lx,cross0/oc0/lc0}{
 \draw[->] (\a.east)--(\b.west);
 \draw[->,dashed] (\a.east)--(\c.west);
}

\node[box,text width=23mm] (spin) at (6.8,-.95)
 {$\mathrm{SPI}_{p}$};
\node[box,text width=23mm] (tagz) at (6.8,-2.15)
 {Obtain $b$ from\\$\mathcal M_{Z,p}$};
\node[box,text width=23mm] (crossn) at (6.8,-3.35)
 {Discard input};

\draw[->] (nonzero.east)--++(.3,0)
 |-node[pos=.74,above] {$u_n$} (spin.west);
\draw[->] (nonzero.east)
 --node[midway,above=2pt] {$t_{Z,n}$} (tagz.west);
\draw[->] (nonzero.east)--++(.3,0)
 |-node[pos=.74,below] {$c_n$} (crossn.west);

\node[lab] (on) at (8.9,-.75)
 {Light: $\mathrm{SPI}_{p}(\rho_n)$};
\node[lab] (ln) at (8.9,-1.15) {$c=\mathrm{untag}$};
\node[lab] (oz) at (8.9,-1.95) {Light: $\rho_0$};
\node[lab] (lz) at (8.9,-2.35) {$c=\mathrm{tagged}_{Z,b}$};
\node[lab] (ocn) at (8.9,-3.15) {Light: $\rho_0$};
\node[lab] (lcn) at (8.9,-3.55) {$c=\mathrm{cross}$};

\foreach \a/\b/\c in {spin/on/ln,tagz/oz/lz,crossn/ocn/lcn}{
 \draw[->] (\a.east)--(\b.west);
 \draw[->,dashed] (\a.east)--(\c.west);
}

\end{tikzpicture}
\caption{Explicit construction of $F$ in Fig.~\ref{fig:modified}.
The input optical system is $B_{\rm in}$, and the optical output
system of each branch is $B_{\rm out}$.
The classical label $c$ is output in system $C_F$.
Solid arrows represent optical systems, and dashed arrows
represent classical labels.
The vacuum state is denoted by $\rho_0$.
The branch probabilities are given by Eqs.~\eqref{eq:untag-n},
\eqref{eq:F-nonvac-weights}, and \eqref{eq:F-vac-weights}.}
\label{fig:F}
\end{figure*}
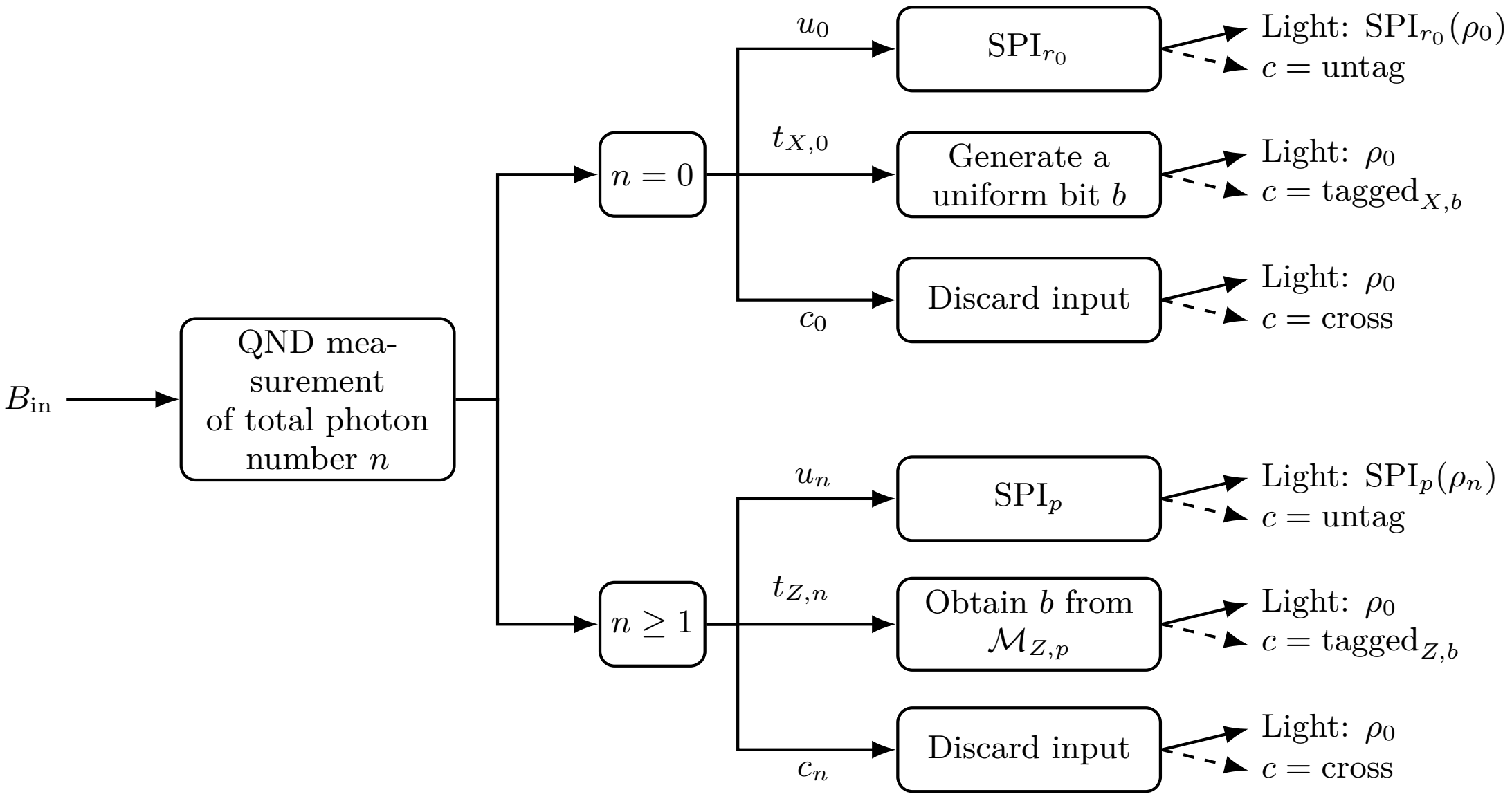

In this subsection, we introduce the single-photon insertion (SPI) channel, which is used to construct $F$ (Fig.~\ref{fig:F}) in the proof of Lemma~\ref{lem:exact}.

This SPI channel can also be used to establish the decomposition mentioned
in the Introduction for active BB84: a measurement with dark counts can be reproduced by a basis-independent CPTP map $F'_p$ followed by a measurement without dark counts (Proposition \ref{prop:dark_count_simulatability}).

Let $\mathcal M_{\beta,p}$ denote a measurement in basis $\beta\in\{Z,X\}$ using two threshold detectors with the unit efficiency and dark count probability $p$.
More precisely, $\mathcal M_{\beta,p}$ is a POVM map which takes an optical state as input and outputs a bit $b\in\{0,1\}$, or the no-detection outcome $\emptyset$.
Each detector independently produces a dark count with the equal probability $p$.
A single click is assigned the bit associated with the detector that clicked, whereas a double click is assigned a uniformly random bit.

Map $\mathcal M_{\beta,p}$ can be viewed as a unit-efficiency detector for active BB84.
For a fixed basis $\beta\in\{Z,X\}$, it can also be viewed as a unit-efficiency detector at the end of the $\beta$ path in passive BB84.
$\mathcal M_{\beta,0}$ particularly describes the ideal detector, with the unit efficiency and zero dark-count probability.

We also define the single-photon insertion (SPI) channel $\mathrm{SPI}_q$, which replaces the input with $\tau_1$ with probability $q\in[0,1]$ and otherwise leaves it unchanged.
Namely, for a subnormalized state $\rho$ ($\rho\ge0$, $\Tr\rho\le1$), channel $\mathrm{SPI}_q$ acts as
\begin{equation}
 \mathrm{SPI}_q(\rho)=(1-q)\rho+q\tau_1\Tr\rho.
 \label{eq:spi-channel}
\end{equation}
Here $\tau_1$ is the maximally mixed state in the single-photon subspace
\begin{equation}
 \tau_1:=\frac{I_2}{2},
 \label{eq:tau1}
\end{equation}
with $I_2$ being the identity operator in the same subspace.

We now show that the measurement $\mathcal M_{\beta,p}$, with dark-count probability $p\ge0$, can be simulated by a basis-independent CPTP map $F'_p$ followed by the ideal measurement $\mathcal M_{\beta,0}$.

\begin{proposition}
\label{prop:dark_count_simulatability}
Let $\Pi_n$ be the projector onto the subspace of total photon number $n$.
Then
\begin{equation}
 F_p'(\rho)=\mathrm{SPI}_{2p-p^2}(\Pi_0\rho\Pi_0)
          +\sum_{n\ge1}\mathrm{SPI}_{p}(\Pi_n\rho\Pi_n)
 \label{eq:Fprime}
\end{equation}
is a CPTP map independent of the basis $\beta$, and
\begin{equation}
 \mathcal M_{\beta,p}=\mathcal M_{\beta,0}\circ F'_p
 \label{eq:active-dark-reduction}
\end{equation}
holds for both bases $\beta\in\{Z,X\}$.
\end{proposition}

\begin{proof}
The POVM elements of $\mathcal M_{\beta,p}$ are block diagonal in total photon number $n$.
Thus we may consider the vacuum and nonvacuum inputs separately.

For a vacuum input, the actual measurement outputs bit $b$ with probability $p(1-p)+p^2/2=(2p-p^2)/2$.
This is reproduced by $F'_p$ sending $\tau_1$ with probability $2p-p^2$, followed by the ideal measurement producing a uniformly random bit.
With the remaining probability $(1-p)^2$, $F'_p$ sends the vacuum and reproduces the no-detection outcome.

For a nonvacuum input, unit detection efficiency ensures that at least one detector clicks.
On the subspace that produces a double click even without dark counts, the assigned bit is already uniform, so replacement with $\tau_1$ preserves its distribution.
On a subspace that produces a single click, a dark count in the other detector occurs with probability $p$, producing a double click and hence a uniformly random bit.
The probability of obtaining the opposite bit is therefore $p/2$.
This is reproduced by replacement with $\tau_1$ with probability $p$. 
The preceding calculations verify Eq.~\eqref{eq:active-dark-reduction}
for inputs with a fixed total photon number.
The positive-operator-valued measure (POVM) elements corresponding
to the two sides of Eq.~\eqref{eq:active-dark-reduction}
are block diagonal in total photon number,
so the equality also holds for arbitrary input states.
The equivalence concerns the processed record consisting of a bit or no detection; it does not require reproducing all raw click patterns.
\end{proof}

\subsection{Main lemma: Equivalence of the actual passive and virtual receivers}\label{sec:main-lemma}

Throughout this subsection, we restrict attention to a single round. Omitting the round index $i$ in the measurement record $Y_i$ defined in Eq.~\eqref{eq:record} of Sec.~\ref{sec:protocol}, we write
\[
 Y\in\{Z0,Z1,X0,X1,\mathrm{cross},\emptyset\}.
\]
Our goal is to reproduce the statistics of the record $Y$ using a CPTP map $F$ applied before basis selection, followed by an ideal active measurement.

The input system $B_{\rm in}$ of $F$ is the light incident on Bob after moving the common loss to the input as described in Sec.~\ref{sec:receiver}.
The outputs are an optical system $B_{\rm out}$ and a system $C_F$ storing a classical label.
The classical label takes values in
\[
 \begin{split}
 c\in\{&\mathrm{untag},\mathrm{tagged}_{Z,0},\mathrm{tagged}_{Z,1},\\
        &\mathrm{tagged}_{X,0},\mathrm{tagged}_{X,1},\mathrm{cross}\}.
 \end{split}
\]
The map can be written as
\begin{equation}
 F(\rho)=\sum_c F_c(\rho)_{B_{\rm out}}
          \otimes\ket c\bra c_{C_F}.
 \label{eq:F-output}
\end{equation}
Each $F_c$ is a completely positive, trace-nonincreasing map, and the probability of branch $c$ is $\Tr F_c(\rho)$.
Together, the branches satisfy $\sum_c\Tr F_c(\rho)=1$.
The optical output can be the vacuum.
Let $R$ denote the systems that may be correlated with the input light, including those held by Alice and Eve and those associated with other transmission rounds.

\begin{mainlemma}[Decomposition of a passive receiver into an ideal active measurement]\label{lem:exact}
Consider the unit-efficiency model with effective BS transmittance $1/2\le T<1$ obtained by moving the common input loss of the receiver in Definition~\ref{def:receiver} into the quantum channel.
There exists a CPTP map $F$, applied before basis selection, from the input optical system $B_{\rm in}$ to the optical system $B_{\rm out}$ and the classical-label system $C_F$ in Eq.~\eqref{eq:F-output}.

For the optical output of $F$, independently select basis $Z$ or $X$ with probability $T$ or $1-T$, respectively, and perform an active measurement with zero dark-count probability and unit efficiency. 
Determine the receiver's record $Y$ as follows:
\begin{enumerate}
\item $c=\mathrm{untag}$: If the active measurement in basis $\beta$ produces bit $b$, set $Y=\beta b$; if there is no detection, set $Y=\emptyset$.
\item $c=\mathrm{tagged}_{\beta,b}$: Set $Y=\beta b$ as specified by $F$.
\item $c=\mathrm{cross}$: Set $Y=\mathrm{cross}$.
\end{enumerate}
For every input state $\rho_{B_{\rm in}R}$, the joint state of the record $Y$ and $R$ generated in this way is identical
to that obtained with the actual passive receiver.
\end{mainlemma}

The virtual receiver is shown in Fig.~\ref{fig:modified}. 
On the tagged and cross branches in Fig.~\ref{fig:modified},
$F$ sends the vacuum state $\rho_0:=\ket{0,0}\bra{0,0}$ 
to the ideal active measurement.
Although this measurement still randomly chooses the $Z$ or $X$ basis,
the receiver ignores that choice and determines the record $Y$
from the classical output $c$ of $F$:
$Y=\beta b$ for $c=\mathrm{tagged}_{\beta,b}$,
and $Y=\mathrm{cross}$ for $c=\mathrm{cross}$.

\begin{table}[!t]\small\centering
\caption{
Probabilities of the measurement records of the actual passive receiver, conditioned on the input photon number \(n\).
The entries for $Zb$ and $Xb$ apply separately to $b=0,1$.
Here $a:=(1-p)^2$ is the probability that neither detector in one basis path produces a dark count. 
For a normalized \(n\)-photon state \(\rho_n\) with \(n\ge1\), \(v_{\beta b}\) is the probability of obtaining bit \(b\) in the measurement \(\mathcal M_{\beta,p}\) (\(\beta\in\{Z,X\}\)).
A single click is assigned the bit associated with the detector, and a double click is assigned a uniformly random bit.
}
\label{tab:target-statistics}
\begin{tabular}{@{}lll@{}}\toprule
Record $Y$ & $n=0$ & $n\ge1$\\\midrule
$Zb$ & $a(1-a)/2$ & $aT^n v_{Zb}$\\
$Xb$ & $a(1-a)/2$ & $a(1-T)^n v_{Xb}$\\
$\mathrm{cross}$ & $(1-a)^2$ & $1-a[T^n+(1-T)^n]$\\
$\emptyset$ & $a^2$ & $0$\\\bottomrule
\end{tabular}
\end{table}

\subsection{Proof of the main lemma: Constructing \texorpdfstring{$F$}{F} from the target statistics}
\label{sec:F-construction}
We prove Lemma~\ref{lem:exact} by constructing $F$ to reproduce the measurement statistics of the actual passive receiver.

\begin{proof}
We construct $F$ along the branches in Fig.~\ref{fig:F}.

The POVM elements of the actual passive receiver are block diagonal
in $n$, so inserting this measurement does not change the measurement statistics.
Thus we insert a quantum nondemolition (QND) measurement of the total input photon number $n$ at the first step of $F$.

Matching the probability of each measurement outcome for every input state with a fixed photon number $n$ establishes equality of the POVM elements
of the actual passive and virtual receivers.
This also establishes equality of the joint state of the record $Y$ and the systems held by Alice and Eve, among others [Eq.~\eqref{eq:receiver-cq-equivalence}]. 
On the untagged branch, apply $\mathrm{SPI}_q$ in Eq.~\eqref{eq:spi-channel} to simulate dark counts and send the resulting light to the ideal active measurement.
We determine the branch probabilities and the replacement probability of $\mathrm{SPI}_q$ by matching the measurement statistics of the actual passive receiver.

\paragraph{Measurement statistics to be reproduced.}
Let $a:=(1-p)^2$ be the probability that neither detector in one basis path produces a dark count.
For a normalized input $\rho_n$ with $n\ge1$, let $v_{\beta b}$ be the probability of obtaining bit $b$ under $\mathcal M_{\beta,p}$. 
Table~\ref{tab:target-statistics} lists the measurement outcome
probabilities of the actual passive receiver for each input
photon number $n$.
We construct $F$ so that the virtual receiver in
Fig.~\ref{fig:modified} reproduces these probabilities.
For example, a nonvacuum input produces detections in the $Z$ path alone only if all photons enter that path and no dark count occurs in the empty $X$ path, giving probability $aT^n$.
Because the BS has the same splitting ratio for both modes, the normalized state is unchanged when all photons enter one path, so $v_{\beta b}$ gives the bit distribution in that path.

\paragraph{Nonvacuum inputs.}
For $T>1/2$ and multiphoton inputs ($n\ge2$), the actual passive receiver's
ratio $T^n:(1-T)^n$ of $Z$-only to $X$-only detection probabilities is more biased toward $Z$ than the ideal active basis-selection ratio $T:(1-T)$. 
We choose the untagged-branch probability to match the actual
passive receiver's $X$-only detection probability.
We then choose the $Z$-tagged-branch probability so that the
untagged and tagged branches together match the actual passive
receiver's $Z$-only detection probability.
Let $u_n$ be the untagged-branch probability.
On this branch, apply $\mathrm{SPI}_p$ and send its output to the ideal active measurement.
The output is nonvacuum, so detection always succeeds, and Eq.~\eqref{eq:active-dark-reduction} gives the bit distribution $v_{\beta b}$ in each basis.
Matching the $X$-detection probability, $(1-T)u_n=a(1-T)^n$, gives
\begin{equation}
 u_n=a(1-T)^{n-1}.
 \label{eq:untag-n}
\end{equation}
The $Z$-detection probability on this branch is $Tu_n$.
A tagged $Z$ branch supplies the difference from the actual passive probability $aT^n$.
Writing its probability as $t_{Z,n}$ and the cross-branch probability
as $c_n$, and using the normalization condition
$u_n+t_{Z,n}+c_n=1$, we obtain
\begin{equation}
\begin{split}
t_{Z,n}&=aT^n-Tu_n,\\
c_n&=1-u_n-t_{Z,n}=1-a[T^n+(1-T)^n].
\end{split}
\label{eq:F-nonvac-weights}
\end{equation}
On the tagged $Z$ branch, measure the original input with $\mathcal M_{Z,p}$, record the resulting bit $b$ as $c=\mathrm{tagged}_{Z,b}$, and output the optical vacuum.
On the cross branch, discard the input, output the vacuum, and record $c=\mathrm{cross}$.
Since $T\ge1/2$, all branch probabilities are nonnegative and sum to one.
The record-generation rules give
\begin{align}
 \Pr[Y=Zb\mid n]&=(Tu_n+t_{Z,n})v_{Zb}
                 =aT^n v_{Zb},\notag\\
 \Pr[Y=Xb\mid n]&=(1-T)u_n v_{Xb}
                 =a(1-T)^n v_{Xb},
 \label{eq:bitwise-match}
\end{align}
reproducing the bit-resolved probabilities in Table~\ref{tab:target-statistics}.

\paragraph{Vacuum inputs.}
For vacuum inputs, the actual passive receiver clicks only because
of dark counts, with equal $Z$-only and $X$-only probabilities $a(1-a)$.
On the untagged branch, these dark counts are simulated by sending
a single photon to the ideal active measurement, which chooses
$Z$ and $X$ with probabilities $T$ and $1-T$.
For $T>1/2$, matching the actual passive receiver's $Z$-only
probability therefore gives too few $X$-only detections.
We add an $X$-tagged branch to $F$ so that the total $X$-only
probability also matches that of the actual passive receiver. 
Let $u_0$ be the probability of the untagged branch.
On this branch, replace the vacuum by the single-photon state
$\tau_1$ with probability $r_0\in[0,1]$, to be determined below.
This operation is $\mathrm{SPI}_{r_0}$, whose output is sent
to the ideal active measurement in Fig.~\ref{fig:modified}.
Matching this branch's $Z$-detection probability to the actual passive receiver gives an $X$-detection probability equal to $(1-T)/T$ times the actual passive probability.
To supply the difference, introduce a branch of probability $t_{X,0}$ that chooses a uniformly random bit $b$ and outputs $c=\mathrm{tagged}_{X,b}$.
Its optical output is the vacuum.
The conditions for matching the $Z$-only, $X$-only, and no-detection probabilities, respectively, are
\begin{equation}
\begin{split}
Tu_0r_0&=a(1-a),\\
(1-T)u_0r_0+t_{X,0}&=a(1-a),\\
u_0(1-r_0)&=a^2.
\end{split}
\label{eq:F-vac-matching}
\end{equation}
Solving these equations and using the normalization condition $c_0=1-u_0-t_{X,0}$ yields
\begin{equation}
\begin{aligned}
u_0&=\frac{a[1-(1-T)a]}{T},&
r_0&=\frac{1-a}{1-(1-T)a},\\
t_{X,0}&=a(1-a)(2-T^{-1}),&c_0&=(1-a)^2.
\end{aligned}
\label{eq:F-vac-weights}
\end{equation}
With the remaining probability $c_0$, record a cross click and output the vacuum.
Again, $u_0+t_{X,0}+c_0=1$, all branch probabilities are nonnegative, and the replacement probability satisfies $0\le r_0\le1$.

Although the ideal active measurement receives the vacuum on the tagged and cross branches, the overall receiver records the tagged bit or a cross click according to the classical label.
The overall no-detection outcome occurs only on the untagged branch of a vacuum input when no replacement with a single photon takes place.
Its probability is $u_0(1-r_0)=a^2$.

\paragraph{Equivalence including the systems held by Alice and Eve.}
The construction reproduces the actual passive probabilities for every input $\rho_n$ in each photon-number subspace.
Moreover, both receivers' POVM elements are block diagonal
in total photon number. 
The POVM elements $E_y$ corresponding to each record $y$ therefore coincide.
For the input light $B_{\rm in}$ and the other systems $R$ defined in Sec.~\ref{sec:main-lemma}, the state after measuring and discarding the light is
\begin{equation}
 \rho_{YR}=\sum_y\ket y\bra y\otimes
 \Tr_{B_{\rm in}}[(E_y\otimes I_R)\rho_{B_{\rm in}R}].
 \label{eq:receiver-cq-equivalence}
\end{equation}
Here $\ket y$ are orthogonal states representing the records $y$, and $I_R$ is the identity operator on $R$.
Equality of the POVM elements implies equality of this joint state.
The main lemma thus holds including the correlations of the input light with the systems held by Alice and Eve and with other transmission rounds.
\end{proof}

\section{Proof of the reduction theorem}\label{sec:proof}

\subsection{Including \texorpdfstring{$F$}{F} in Eve's attack}
\label{sec:attack-reduction}

Let $\mathcal E$ be an arbitrary attack on the actual passive protocol, including the effect of nonunit detector efficiency.
By Lemma~\ref{lem:exact}, we can replace the receiver with $F$ followed by ideal active measurements.

If we pessimistically regard the map $F$ for each round as part of Eve's attack, 
we can view 
\begin{equation}
 \mathcal E'=F^{\otimes N}\circ\mathcal E
 \label{eq:composed-attack}
\end{equation}
as an attack in the bridging protocol that sends the optical outputs to Bob and retains the classical flags produced by $F$.
Following this attack, Bob chooses his basis independently after $F$, with $p_Z^B=T$.

This entire process can be viewed as tag generation by $F$ followed by Bob's measurement of the untagged part as in active BB84. It is therefore equivalent to the bridging protocol.

The optical outputs of the tagged and cross branches are the vacuum state, so only untagged branches can produce detections in the ideal active measurement.
For Bob's tagged bits, we use the bits specified by $F$; for Alice's tagged bits, we use her bits from the corresponding transmission rounds.
Inserting the bits from basis-matched tagged rounds in transmission order reproduces the $Z$-basis sifted key and the $X$-basis sample of the actual passive protocol.
To compare with the actual output key length, we define the value of the bridging key-length formula evaluated at the true tag count $\Ntag$ as
\begin{equation}
\begin{split}
L_{\rm true}:=\mathcal L_{\rm bri}\big(
&N,(p_Z^A,T),\Nsift,\Ntag,\\
&\NerrObs,\lambda_{\rm EC};\eps\big).
\end{split}
\label{eq:oracle-key}
\end{equation}
By Assumption~\ref{ass:active-formula}, the bridging protocol is $\eps$-secure whenever its output length does not exceed $L_{\rm true}$.

\subsection{Security of the key length obtained by substitution}
\label{sec:length-transfer}

The actual passive protocol does not know the true tag count $\Ntag$.
It therefore substitutes the estimated upper bound $\NtagU$ for the tag-count argument of $\mathcal L_{\rm bri}$ and chooses the output key length $\ell_{\pas}$ no larger than the resulting value [Eq.~\eqref{eq:key-transfer}].
On the event $G:=\{\Ntag\le\NtagU\}$ that tag estimation succeeds, we have $\Nsift-\NtagU\le\Nsift-\Ntag$.
Applying the nondecreasing dependence on the sifted-key length in condition~1 of Assumption~\ref{ass:active-formula} to Eqs.~\eqref{eq:normal-bridge-BB84KL_relation} and~\eqref{eq:key-transfer} gives
\begin{equation}
 \ell_{\pas}\le L_{\rm true}
 \label{eq:length-order}
\end{equation}
whenever a key is output.

In the virtual attack $\mathcal E'$, Eve retains the classical output of $F$ as an internal record.
Since $G$ can be determined from this output and the public information, the terms associated with the classical records can be grouped according to $G$ and $G^c$.
The final state in the virtual description can thus be written as $\rho=\rho_G+\rho_{G^c}$, where the two parts are subnormalized, with $\Tr\rho_G=\Pr[G]$ and $\Tr\rho_{G^c}=\Pr[G^c]$.
Replacing the key with an ideal key preserves the classical records, so the ideal state admits the same decomposition,
$\rho^{\rm ideal}=\rho_G^{\rm ideal}+\rho_{G^c}^{\rm ideal}$.

For the security proof, consider aborting on $G^c$.
Every execution that outputs a key then satisfies $\ell_{\pas}\le L_{\rm true}$.
The security guarantee of the bridging formula bounds the contribution from successful tag estimation as $d(\rho_G,\rho_G^{\rm ideal})\le\eps$.
For failed estimation, the trace distance is at most $\Pr[G^c]$, the common trace of the real and ideal parts.
Therefore,
\[
\begin{aligned}
d(\rho,\rho^{\rm ideal})
&\le d(\rho_G,\rho_G^{\rm ideal})
   +d(\rho_{G^c},\rho_{G^c}^{\rm ideal})\\
&\le\eps+\Pr[G^c]
 \le\eps+\delta_{\tagg}.
\end{aligned}
\]
Discarding the tag records retained internally in the virtual description yields the output state of the actual passive protocol and cannot increase the trace distance.
This proves Theorem~\ref{thm:transfer}.

\section{\texorpdfstring{Discussion}{Discussion}}
\label{sec:conclusion}
We have exactly decomposed a passive BB84 receiver with an unbalanced BS and dark counts into a basis-independent quantum channel, ideal active measurements, and tagged classical outputs (Lemma \ref{lem:exact}).
By using this decomposition we showed that existing active security proofs and key-length formulas can be reused to obtain the passive key length (Theorem \ref{thm:transfer}), under the applicability conditions in Assumption~\ref{ass:active-formula}.

The receiver equivalence established here holds independently of Alice's source.
For the key-length formula, we considered BB84 with an ideal single-photon source and showed how to obtain a passive formula from an existing active formula.
The same receiver decomposition can also be used in more realistic settings, including decoy-state BB84~\cite{LoMaChen2005} and imperfect sources.
Connecting the decomposition to active security proofs for these settings and explicitly deriving the corresponding passive key-length formulas remain topics for future work.

\begin{acknowledgments}

The mathematical ideas underlying this work emerged in the course of commissioned technical activities associated with Quantum Forum, Japan. 
We thank Masahide Sasaki for his support of these activities. We also thank Kiyoshi Tamaki, Ryutaroh Matsumoto, and Toshimori Honjo for their careful reviews and constructive scientific discussions in the course of these activities.

ChatGPT (OpenAI; [Astra/6]) was used to assist with
revising the wording and presentation of scientific explanations in the manuscript. The authors directed the revisions, reviewed and edited the
resulting text, and take full responsibility for the content
of this work.
\end{acknowledgments}

\section*{Data availability}
No data were created or analyzed in this study.

\appendix

\section{Real and ideal states with variable key lengths}
\label{app:security}
We specify the states used in the security definition of Sec.~\ref{sec:security-definition}.
For a public record $c$ indicating success, let $\ell(c)$ be the key length and define
\begin{equation}
 \tau^{(\ell)}_{S_AS_B}
 =2^{-\ell}\sum_{z\in\{0,1\}^{\ell}}
             \ket{z,z}\bra{z,z}.
 \label{eq:ideal-key}
\end{equation}
On abort, we set $\ell(c)=0$ and $\tau^{(0)}=\ket{\bot,\bot}\bra{\bot,\bot}$.
The real and ideal final states are
\begin{align}
\rho_{S_AS_BEC}
&=\sum_c p_c\,\ket c\bra c_C\otimes\rho^c_{S_AS_BE},
\label{eq:real-state}\\
\rho^{\rm ideal}_{S_AS_BEC}
&=\sum_c p_c\,\ket c\bra c_C\otimes
\tau^{(\ell(c))}_{S_AS_B}\otimes\rho_E^c,
\label{eq:ideal-state}
\end{align}
where $\rho_E^c=\Tr_{S_AS_B}\rho^c_{S_AS_BE}$.
Here $p_c$ is the probability of the public record $c$, and each $\rho^c$ is the normalized state conditioned on that record.

If the probability of unequal keys satisfies $\Pr[S_A\ne S_B,\mathrm{success}]\le\eps_{\rm cor}$ and the secrecy error of Alice's key satisfies $d(\rho_{S_AEC},\rho^{\rm ideal}_{S_AEC})\le\eps_{\rm sec}$, the overall security error is at most $\eps_{\rm cor}+\eps_{\rm sec}$~\cite{PortmannRenner}.

\section{Verification of Assumption~\ref{ass:active-formula} for established proof approaches}
\label{sec:well-known-methods}
We show that key-length formulas obtained using established security-proof approaches---specifically, the Shor--Preskill approach~\cite{ShorPreskill}, Koashi's approach~\cite{Koashi,HayashiTsurumaru}, and Renner's approach~\cite{RennerThesis,Tomamichel}---satisfy Assumption~\ref{ass:active-formula} when universal$_2$ hashing is used for privacy amplification. 
Condition~1 is straightforward to verify, so we prove condition~2. 

In the bridging protocol, the tagged part is assigned no secrecy.
Privacy amplification is applied to the entire sifted key, with $\Ntag$ added to the number of bits removed for the untagged part.
Since secrecy is evaluated for the untagged part, we first establish security using the untagged key length $\Nunt$ and the untagged sample error count $\NerrU$ as inputs to $\Gamma$.
Specifically, we show that the bridging protocol is $\eps$-secure at the key length obtained by replacing $\NerrObs$ with $\NerrU$ in Eq.~\eqref{eq:normal-bridge-BB84KL_relation}.

Once this guarantee is established, $\NerrObs\ge\NerrU$ [Eq.~\eqref{eq:sample-addition}] and the monotonicity of the key length in the error count imply 
that the bridging protocol is $\eps$-secure with key length $\ell_{\rm bri}$ satisfying Eq.~\eqref{eq:normal-bridge-BB84KL_relation}.
We therefore show, in turn for the Shor--Preskill approach, Koashi's approach, and Renner's approach, that the additional privacy amplification for tagged bits guarantees the key length evaluated using the untagged quantities.

In each proof, $\lambda_{\rm EC}$ denotes an upper bound on public leakage from error correction and error verification on the entire sifted key, including tagged bits. 
The original active security proof allows Eve to retain arbitrary ancillary systems, including those used to generate the tags. We apply its security guarantee with these systems held by Eve.

\subsection{Shor--Preskill approach}\label{app:Shor-Preskill}

In the Shor--Preskill approach, we use the tagging argument in which all phase information of the tagged qubits is treated as unknown~\cite{ShorPreskill,GLLP}.
Let $A$ be an upper bound on the number of candidate phase-error strings for the untagged part, obtained from the original active proof using the untagged key length $\Nunt$ and the untagged sample error count $\NerrU$.
After tag insertion, the number of candidates is at most $2^{\Ntag}A$.

In standard phase-error correction with uniformly random linear checks, the decoding failure probability with $m$ check bits is bounded by $A2^{-m}$.
Adding $\Ntag$ check bits preserves the original bound after tag insertion because
\[
 (2^{\Ntag}A)\,2^{-(m+\Ntag)}=A2^{-m}.
\]
The original bound also applies to the probability that the untagged phase-error string lies outside the candidate set. 
Thus, removing $\Ntag$ additional bits by privacy amplification preserves the original upper bounds on the failure probabilities of phase-error estimation and phase-error correction. 
This argument uses the correspondence between linear privacy amplification and virtual phase-error correction, together with the bound on the failure probability of the checks used in the original proof~\cite{HayashiTsurumaru,TsurumaruUnified}.

An analogous argument applies when universal$_2$ hash functions are used for privacy amplification. This follows from the known fact that universal$_2$ hash functions are also 2-almost dual universal$_2$ \cite{TsurumaruHayashi2013}. Consequently, the failure probability of the corresponding (virtual) phase-error correction can be bounded by twice that for uniformly random linear checks.

\subsection{Koashi's approach}

Koashi's approach narrows down the candidates for the virtual $X$-measurement outcomes using random parity checks~\cite{Koashi,HayashiTsurumaru}.
Even when all $X$-measurement outcomes in the tagged part are treated as unknown, the number of candidates increases by at most a factor of $2^{\Ntag}$. 
The bound in Appendix~\ref{app:Shor-Preskill} implies that adding $\Ntag$ random parity checks preserves the original upper bound on the probability of failing to identify the virtual $X$-measurement outcomes. 
Applying the original complementarity-based security argument therefore gives the same security error when $\Ntag$ additional bits are removed by privacy amplification.

An analogous argument applies when universal$_2$ hash functions are used for privacy amplification, due to the same argument as in the last paragraph of Appendix \ref{app:Shor-Preskill}.

\subsection{Renner's approach}\label{app:entropy}

We show that the lower bound on the smooth min-entropy of the untagged key supplied by the original active proof also applies to the entire sifted key after tag insertion.
The inputs to the original proof are the untagged key length $\Nunt$ and the untagged sample error count $\NerrU$.

Let $E_0$ be Eve's system before tag generation and $D$ the information already made public.
For the passive protocol, $E_0$ includes the classical flags produced by $F$ and Alice's retained qubits in tagged rounds.
We apply the original entropy lower bound for the untagged key before error correction relative to this enlarged information $E_0D$.

Eve processes $(E_0,D)$ to generate the tag record $D_{\tagg}$ and retains the residual system $E_1$.
For both the $Z$ sifted key and the $X$ sample, $D_{\tagg}$ contains the tag values and the insertion positions after sifting.
In particular, $J,W_A,W_B$ for the $Z$ sifted key can be recovered from $D_{\tagg}$.
Measurements on the tag systems commute with measurements on the untagged systems, so we may evaluate security in the order in which the untagged strings are obtained first and the tags are generated afterward.

The min-entropies before and after tag insertion satisfy
\begin{align}
 H_{\min}(K_A\mid E_1DD_{\tagg})
 &=H_{\min}(K_A^{\unt}\mid E_1DD_{\tagg})
 \notag\\
 &\ge H_{\min}(K_A^{\unt}\mid E_0D).
 \label{eq:tag-entropy}
\end{align}
The equality holds because, given $J,W_A$, inserting or removing tags uniquely converts $K_A^{\unt}$ into $K_A$ and vice versa.
The inequality follows from data processing: processing her own information cannot improve Eve's ability to guess the untagged key~\cite{RennerThesis,TomamichelThesis}.
Treating insertion as an isometry on the classical support gives the same relations for smooth min-entropy.

The original entropy lower bound therefore remains valid after tag insertion.
Subtracting the public leakage $\lambda_{\rm EC}$ from error correction and error verification on the entire string, including tags, and applying the leftover hashing lemma~\cite{LHL} with the same smoothing and privacy-amplification error parameters as in the original proof guarantees $\eps$-security of the bridging protocol up to the key length obtained by substituting $\Nunt,\NerrU$ into Eq.~\eqref{eq:normalBB84-KL}. 
After tag insertion, PA is applied to the entire sifted key,
which is $\Ntag$ bits longer than the original untagged key.
The entropy lower bound remains unchanged.
It is therefore sufficient to remove $\Ntag$ more bits by PA than required for the untagged key.

\section{Bounding the number of tags from cross clicks}
\label{app:statistics}

For the protocol in Sec.~\ref{sec:protocol}, we bound the number of tags
$\Ntag$ in the sifted key using the cross-click count $\Ncross$
among the rounds in which Alice selects the $Z$ basis.
Throughout this appendix, a tag means a $Z$ tag.
For $T=1/2$, no $Z$ tags occur, so we set $\NtagU=0$.
Below, we assume $1/2<T<1$. 
Before quantum communication begins, we fix $\delta_{\tagg}\in(0,1)$
as an upper bound on the failure probability of tag-count estimation.

\subsection{Relation between tag and cross-click probabilities}
\label{app:ratio}

Let $\kappa:=(2T-1)/[2(1-T)]$.
Substituting Eq.~\eqref{eq:untag-n} into Eq.~\eqref{eq:F-nonvac-weights}
gives, for $n\ge1$,
\begin{equation}
 2(1-T)(\kappa c_n-t_{Z,n})
 =(2T-1)-a[T^n-(1-T)^n]\ge0.
 \label{eq:ratio}
\end{equation}
The inequality follows from $0<a\le1$ and
$0\le T^n-(1-T)^n\le2T-1$.
The difference $T^n-(1-T)^n$ equals $2T-1$ at $n=1$
and is nonincreasing in $n$.
For a vacuum input, $t_{Z,0}=0$, so
$t_{Z,n}\le\kappa c_n$ also holds for $n=0$.
Using this inequality, we obtain a lower bound on the
probability of a cross click given that the round produced
either a $Z$ tag or a cross click:
\begin{equation}
 q_n:=\frac{c_n}{c_n+t_{Z,n}}
 \ge\frac{1}{1+\kappa}
 =2(1-T)=:r.
 \label{eq:cross-fraction}
\end{equation}
If the denominator is zero, neither a tag nor a cross click occurs.
The lower bound $r$ in Eq.~\eqref{eq:cross-fraction} is independent of
the incoming photon number $n$ and the dark-count probability $p$.

\subsection{Bounding the tag count from the observed cross-click count}
\label{app:binomial}

The inputs to the calculation are $T$, $\delta_{\tagg}$,
and the observed counts $\Naz$, $\Ncross$, and $\Nsift$.
We first compare the cross-click count with a binomial random variable
and then use this comparison to define the upper bound.

\paragraph{Comparison with a binomial distribution.}
For the proof, fix Alice's basis sequence, the incoming photon-number
sequence obtained by the virtual QND measurement, and the set of
positions at which Alice selects $Z$ and either a tag or a cross click occurs.
Let $L$ be the size of this set, so that 
\[
L=\Ntag+\Ncross.
\]
The comparison below and the proof of the lemma use probabilities
conditioned on these quantities.

Given the input photon numbers, $F$ chooses a branch independently
in each round using only that round's photon number.
This independence is preserved when we condition on which rounds
produce either a tag or a cross click.

Consider a modified trial at each position: leave a tag unchanged,
and change a cross click to a tag with probability $(q_n-r)/q_n$.
The random choices used for these changes are also independent across positions. 
In the modified trial, the probability of a tag is
\begin{equation}
 (1-q_n)+q_n\frac{q_n-r}{q_n}=1-r.
 \label{eq:modified-tag-probability}
\end{equation}
By Eq.~\eqref{eq:modified-tag-probability}, the cross-click probability
after the change is the same value $r$
at every position.
The resulting cross-click count $K$ therefore follows the binomial
distribution $\Bin(L,r)$ with $L$ trials and success probability $r$.
Moreover, $K\le\Ncross$ always holds because the modification only
changes cross clicks to tags. 
Using this inequality, we obtain, for every integer $k$ with $0\le k\le L$,
\begin{equation}
 \Pr[\Ncross\le k]
 \le\Pr[K\le k]
 =\Pr[\Bin(L,r)\le k].
 \label{eq:cross-binomial-domination}
\end{equation}
We denote the binomial lower-tail probability on the right by $B_r(L,k)$.

\paragraph{Calculating the upper bound on the tag count.}
Using the observed values of $\Ncross$ and $\Naz$,
consider candidate tag counts $j$ in order from $0$ to $\Naz-\Ncross$.
As $j$ increases, $B_r(\Ncross+j,\Ncross)$ decreases. 
If this probability is at most $\delta_{\tagg}$ for some candidate
tag count $j$, the upper bound is one less than the smallest such $j$:
\begin{align}
U_{\tagg}=\min\big\{j:\ &
j\in\mathbb Z,\quad 0\le j\le\Naz-\Ncross,
\notag\\
&B_r(\Ncross+j,\Ncross)\le\delta_{\tagg}\big\}-1.
\label{eq:binomial-upper}
\end{align}
If the probability remains above $\delta_{\tagg}$ even at the largest
candidate, namely if $B_r(\Naz,\Ncross)>\delta_{\tagg}$,
set $U_{\tagg}:=\Naz-\Ncross$.
In either case, define the final upper bound as
\begin{equation}
 \NtagU:=\min\{\Nsift,U_{\tagg}\}.
 \label{eq:tag-bound-binomial}
\end{equation}
We note that this construction and the failure-probability bound below follow the same idea as the Bernoulli-sampling argument of Kawakami, Sasaki, and Koashi \cite[Sec.~III B, Eqs.~(18)--(23)]{BernoulliSampling}. Here we apply that idea to tag estimation using the binomial comparison in Eq.~\eqref{eq:cross-binomial-domination}.

In the following lemma, $U_{\tagg}$ and $\NtagU$ are also treated as
random variables because they depend on the observed data.

\begin{lemma}[Binomial upper bound on the tag count]
\label{lem:tag-bound}
For the protocol in Sec.~\ref{sec:protocol},
the upper bound $\NtagU$ defined in Eq.~\eqref{eq:tag-bound-binomial}
satisfies $\Pr[\Ntag>\NtagU]\le\delta_{\tagg}$.
\end{lemma}

\begin{proof}
The true tag count satisfies $0\le\Ntag\le\Naz-\Ncross$.
By the definition of $U_{\tagg}$ in Eq.~\eqref{eq:binomial-upper}
and the fact that $B_r(\Ncross+j,\Ncross)$ is nonincreasing in $j$,
$\Ntag>U_{\tagg}$ implies
\[
 B_r(\Ncross+\Ntag,\Ncross)
 =B_r(L,\Ncross)
 \le\delta_{\tagg}.
\]
Since $B_r(L,k)$ is nondecreasing in $k$,
if $B_r(L,0)>\delta_{\tagg}$, no cross-click count can satisfy
$B_r(L,\Ncross)\le\delta_{\tagg}$, and hence
$\Pr[\Ntag>U_{\tagg}]=0$.
Otherwise, define
\[
 k_\delta:=
 \max\bigl\{k\in\{0,\ldots,L\}:
 B_r(L,k)\le\delta_{\tagg}\bigr\}.
\]
The value of $k_\delta$ is determined by the fixed $L$ and $\delta_{\tagg}$.
By the same monotonicity,
\[
 B_r(L,\Ncross)\le\delta_{\tagg}
 \quad\Longleftrightarrow\quad
 \Ncross\le k_\delta.
\]
Substituting $k=k_\delta$ into Eq.~\eqref{eq:cross-binomial-domination}
therefore gives
\[
\begin{aligned}
 \Pr[\Ntag>U_{\tagg}]
 &\le
 \Pr[B_r(L,\Ncross)\le\delta_{\tagg}]\\
 &=\Pr[\Ncross\le k_\delta]\\
 &\overset{\eqref{eq:cross-binomial-domination}}{\le}
 B_r(L,k_\delta)\\
 &\le\delta_{\tagg}.
\end{aligned}
\]
The last inequality follows from the definition of $k_\delta$.

The bound is preserved when averaged over the records fixed in the
binomial comparison.
Finally, $\Ntag\le\Nsift$ and Eq.~\eqref{eq:tag-bound-binomial} imply
\[
 \Pr[\Ntag>\NtagU]
 =\Pr[\Ntag>U_{\tagg}]
 \le\delta_{\tagg}.
\]
\end{proof}

\paragraph{Closed-form upper bound.}
In this paragraph, we show that the closed form $\NtagU^{\rm exmp}$ given in Eq.~\eqref{eq:tag-bound-closed} 
indeed provides an upper bound on $\Ntag$. 
For this purpose, we show that $\NtagU^{\rm exmp}$ is at least as large as
$\min\{\Nsift,U_{\tagg}\}$ in Eq.~\eqref{eq:tag-bound-binomial}.
Assume $1/2<T<1$ and use $r$ and $s_{\tagg}$ as defined in
Sec.~\ref{sec:main}.

For integers $L\ge1$ and $k$ satisfying $0\le k\le rL$,
the Chernoff bound for a binomial distribution gives
\[
 B_r(L,k)
 \le
 \exp\!\left[-\frac{(rL-k)^2}{2rL}\right].
\]
Substitute $L=k+j$ into this bound.
Since $s_{\tagg}=\ln(1/\delta_{\tagg})$, the condition
\[
 [r(k+j)-k]^2>2s_{\tagg}r(k+j)
\]
implies $B_r(k+j,k)<\delta_{\tagg}$ in the range $r(k+j)>k$.
Solving this quadratic inequality for $j$ yields, for nonnegative
integers $k$ and $j$,
\[
\begin{gathered}
 j>\frac{(1-r)k+s_{\tagg}
 +\sqrt{s_{\tagg}^{\,2}+2s_{\tagg}k}}{r}\\
 \Rightarrow\quad B_r(k+j,k)<\delta_{\tagg}.
\end{gathered}
\]

When the observed cross-click count is $k$,
Eq.~\eqref{eq:binomial-upper} excludes every candidate tag count
above this threshold.
Thus $\min\{\Nsift,U_{\tagg}\}\le\NtagU^{\rm exmp}$, and
Lemma~\ref{lem:tag-bound} gives
\[
 \Pr[\Ntag>\NtagU^{\rm exmp}]
 \le
 \Pr[\Ntag>\min\{\Nsift,U_{\tagg}\}]
 \le\delta_{\tagg}.
\]
For $T=1/2$, the true tag count is zero, so the same conclusion holds.

\section{An explicit key-length formula based on Bernoulli sampling}\label{app:example}
This appendix gives an example in which the phase-error bound from
Bernoulli sampling \cite[Sec.~III B, Eqs.~(18)--(24)]{BernoulliSampling}
is used to obtain a key length after tag insertion from the secret-key
bound of the original active protocol.
We use the transmission and announcement procedures in Sec.~\ref{sec:protocol}
and the virtual extension of the active protocol in Sec.~\ref{sec:active-input}.
We first give the phase-error bound and the key-length formula,
and then check the applicability conditions.

\subsection{Bounding phase errors from observed errors}\label{app:phase-bound}
Conditioned on Alice and Bob choosing the same basis, let the
probability that the round is assigned to $X$ be
\begin{equation}
 \pi_X=\frac{(1-p_Z^A)(1-p_Z^B)}
 {p_Z^A p_Z^B+(1-p_Z^A)(1-p_Z^B)}\in(0,1).
 \label{eq:bernoulli-probability}
\end{equation}
Fix the phase-error estimation failure probability
$\delta_{\ph}\in(0,1)$ in advance and define
\begin{equation}
\begin{split}
f_{\ph}(\Nerr)=\min\big\{m\in\mathbb Z_{\ge0}:\qquad&\\
\Pr[\Bin(m,\pi_X)\le\Nerr]\le\delta_{\ph}\big\}
&-\Nerr-1.
\end{split}
\label{eq:bernoulli-phase-bound}
\end{equation}
Here $m$ runs over candidate total error counts, and the dependence
of $f_{\ph}$ on $\pi_X$ and $\delta_{\ph}$ is suppressed.
The phase-error count $N_{\ph}$ in the untagged key then satisfies
\begin{equation}
 \Pr[N_{\ph}>f_{\ph}(\NerrU)]\le\delta_{\ph}.
 \label{eq:phase-coverage}
\end{equation}

To obtain this bound, consider the error positions that would arise
if all untagged systems were measured in the $X$ basis.
For a fixed total number of errors at basis-matched positions,
independent basis choices assign errors to the sample according to
a binomial distribution with success probability $\pi_X$;
the remaining errors are phase errors in the key.
Equation~\eqref{eq:bernoulli-phase-bound} excludes candidate total
error counts for which this lower-tail probability is at most $\delta_{\ph}$.
The argument does not require independent and identically distributed
quantum states or conditioning on a fixed sample size.

The bound applies to an arbitrary state of the untagged systems
that is independent of the basis choices, including states correlated
with auxiliary systems used to generate tags.
The secret-key bound obtained from this phase-error estimate can
therefore include these auxiliary systems in Eve's information.

\subsection{Active and passive key-length formulas}\label{app:example-formula}
Use universal$_2$ hashing for PA, with error parameter $\eps_{\PA}$,
and let $\eps_{\rm cor}$ be the correctness error of error verification.
Choose the public leakage bound $\lambda_{\rm EC}$ so that, for each
fixed public record before error correction, the number of possible
subsequent public communication records, including error verification,
is at most $2^{\lambda_{\rm EC}}$.
Using the truncated binary entropy $\overline h$ defined below,
we obtain, for $\Nunt>0$,
\begin{align}
\mathcal L_{\act}
=
&\Nunt\left[1-\overline{h}\!\left(
\frac{f_{\ph}(\Nerr)}{\Nunt}\right)\right]\nonumber\\
&\quad-\lambda_{\rm EC}
-2\log_2\frac{2}{\eps_{\PA}}
\label{eq:example-active}
\end{align}
\cite[Eqs.~(10) and (24)]{BernoulliSampling}.

Here $h_2(x):=-x\log_2x-(1-x)\log_2(1-x)$, and
\[
\bar h(x):=
\begin{cases}
h_2(x),&0\le x\le \tfrac12,\\
1,&x>\tfrac12.
\end{cases}
\]

The function $f_{\ph}(\Nerr)$ in Eq.~\eqref{eq:bernoulli-phase-bound}
is nondecreasing in $\Nerr$.
The properties of the truncated binary entropy $\overline h$ also imply that,
in this example,
$\Nunt-\Gamma(N,\boldsymbol p_Z,\Nunt,\Nerr;\eps)$
is nondecreasing in $\Nunt$ and nonincreasing in $\Nerr$
when the other arguments are fixed.
It therefore satisfies both monotonicity requirements in condition~1
of Assumption~\ref{ass:active-formula}.
Subtracting a fixed $\lambda_{\rm EC}$, rounding down to an integer,
and taking the maximum with zero preserve these monotonicities.
For $\Nunt=0$, the output length is set to zero.
The corresponding security error is
\begin{equation}
 \eps=\eps_{\rm cor}+2\sqrt{\delta_{\ph}}+\eps_{\PA}.
 \label{eq:example-active-security}
\end{equation}

The correspondence with Ref.~\cite{BernoulliSampling} is as follows.
Identify its estimation failure probability with $\delta_{\ph}$
and its PA parameter with $\eps_{\PA}^2/2$ in our notation.
The finite-key term in Eq.~(24) of that reference then becomes
$\log_2[2/(\eps_{\PA}^2/2)]=2\log_2(2/\eps_{\PA})$.
The secrecy error given by its Eq.~(10) is
$\sqrt{2\delta_{\ph}+\eps_{\PA}^2}
\le 2\sqrt{\delta_{\ph}}+\eps_{\PA}$;
adding the correctness error gives Eq.~\eqref{eq:example-active-security}.
This estimate uses the failure probability over the entire protocol
execution in Eq.~\eqref{eq:phase-coverage} and applies to the key length
chosen from the observed record.
In this example,
$\Gamma(N,\boldsymbol p_Z,\Nunt,\Nerr;\eps)
=\Nunt\,\overline{h}\!\left(\frac{f_{\ph}(\Nerr)}{\Nunt}\right)+2\log_2(2/\eps_{\PA})$.

Tags contribute no secret key, and the public leakage is subtracted
for the full string, including the tags.

Following the main theorem, substitute
$\boldsymbol p_Z=(p_Z^A,T)$,
$\Nunt=\Nsift-\NtagU$, and $\Nerr=\NerrObs$.
The resulting key for the actual passive protocol is secure with error
\begin{equation}
\eps_{\rm cor}+2\sqrt{\delta_{\ph}}+\eps_{\PA}+\delta_{\tagg}.
 \label{eq:example-passive-security}
\end{equation}


\begin{thebibliography}{99}
\bibitem{NoteTimeBinEfficiency}
For example, consider a time-bin protocol in which the $Z$ path directly measures
pulse arrival times and the $X$ path uses an unbalanced Mach--Zehnder interferometer
with 50:50 BSs (or couplers).
If only the central interference time window at each output port is accepted,
while the earlier and later noninterfering windows are discarded,
this selection removes half the optical intensity.
If the detector quantum efficiencies and all other losses are equal in the two paths,
then $\eta_X=\eta_Z/2$, or equivalently $\eta_Z=2\eta_X$.
This time-bin protocol thus provides an example with $\eta_Z\neq\eta_X$.
\bibitem{NoteEffectiveEfficiency}
If all detectors have the same efficiency, then $T=T_{\rm phys}$.
The key-length calculation requires the effective transmittance $T$.
By Eq.~\eqref{eq:effective-T}, $T$ can be determined from $T_{\rm phys}$
and the efficiency ratio $\eta_Z/\eta_X$.
Neither the absolute efficiencies $\eta_Z,\eta_X$ nor the common input loss
$\eta_{\rm in}$ need to be known.
The unit-efficiency model used below therefore does not require unit detection
efficiency in the implementation.
In both descriptions, dark counts are applied as an independent process after photon detection.
Moving the loss to the input uses the equality of the efficiencies within each basis
and the specified mode-independent loss model.
\end{thebibliography}
\end{document}